\RequirePackage[hyphens]{url}
\documentclass[a4paper]{article}

\usepackage{graphicx}
\usepackage{amssymb}
\usepackage{amsmath}
\usepackage{hyperref} 
\usepackage{longtable}
\usepackage{xcolor}
\usepackage{float}
\usepackage{subfig}
\usepackage[margin=1cm]{caption}

\usepackage{cite}
\usepackage[margin=1.178in]{geometry}
\usepackage[affil-it]{authblk}

\begin{document}

\title{An Aerodynamic Analysis of Recent FIFA World Cup Balls}

\author{Adrian L. Kiratidis\thanks{Email: \texttt{adriankiratidis@gmail.com}; Corresponding author}\,\, and Derek B. Leinweber\thanks{Email: \texttt{derek.leinweber@adelaide.edu.au}}}
\affil{Special Research Centre for the Subatomic Structure of Matter, Department of Physics, The University of Adelaide, SA, 5005, Australia.}
\maketitle
\begin{abstract}
Drag and lift coefficients of recent FIFA world cup balls are examined.  We fit a novel functional form to drag coefficient curves and in the absence of empirical data provide estimates of lift coefficient behaviour via a consideration of the physics of the boundary layer.  Differences in both these coefficients for recent balls, which result from surface texture modification, can significantly alter trajectories.  Numerical simulations are used to quantify the effect these changes have on the flight paths of various balls.  Altitude  and temperature variations at recent world cup events are also discussed.  We conclude by quantifying the influence these variations have on the three most recent world cup balls, the Brazuca, the Jabulani and the Teamgeist.  While our paper presents findings of interest to the professional sports scientist, it remains accessible to students at the undergraduate level.
\end{abstract}

\section{Introduction}
The aerodynamics of various sports balls has been an area of long-standing interest for both the general public~\cite{TubulentTimes, SurpriseAttack, Swingersdelight, BehindtheSeams, CricketInFullSwing} and the professional sports scientist~\cite{MehtaRD, MehtaRDandPallisJM, HaakeSJGoodwillSRandCarreMJ}.  Golf balls~\cite{Alam-Golf, Nauro-Golf}, cricket balls~\cite{Mehta-Cricketball, Sayers-Cricketball}, baseballs~\cite{BaseballPaper, BaseballRef} and spheres more generally~\cite{AchenbachEhighRe, AchenbachEroughsurf} have all been areas of research interest.  
Of all the sports balls, there is more worldwide interest in the aerodynamics of football (soccer) balls than any other.  This interest typically peaks around the time of the FIFA World Cup and it is not hard to see why.

The FIFA World Cup is perhaps the most watched sporting event globally.  According to FIFA, more than 3.2 billion people, or almost half the world's  population watched at least part of the 2010 incarnation~\cite{fifa-figures}, which netted FIFA \$885 million US dollars in broadcasting and marketing rights~\cite{fifa-report}.  The 2014 World Cup also broke numerous viewing records in a number of different regions~\cite{fifa-report-brasil}.  With the eyes of the media world on the competition there is never a shortage of controversy, both on and off the field.  One such topic that has caused controversy in recent times is that of the ball, where the aerodynamic properties of the surface of the ball has played a controversial role. 

In 1978 the unmistakable `Tango' design was developed for the world cup in Argentina.  The ball was made of leather with a waterproofing coating, and had 20 panels with the visually iconic `triads' creating the impression of 12 circles.  A Tango type ball was used until 2002, when the ball manufacturers, Adidas, opted for a change.  The 32 Panel Adidas Fevernova, with its thick polyurethane surface which included a layer of purpose built foam was designed.  It was built for a `more precise and predictable flight path', and yet the controversy came.  Legendary Italian shot-stopper Gianluigi Buffon labelled the Fevernova ``a ridiculous kiddy's bouncing ball'', while the Brazilian star Rivaldo told reporters the ball soars too far when kicked.

For the 2006 FIFA world cup in Germany the ball was the 14 Panel Teamgeist.  Its thermally bonded panels represented a rather radical change when compared to the stitched panels seen in previous balls.  The Teamgeist was the roundest ball to date, had very smooth panels and was essentially water-proof.  It was expected to perform more uniformly and predictably, but once again the complaints came.  Germany's keeper Oliver Kahn said the ball was ``built in favour of the strikers", while Brazilian Roberto Carlos, who has his fair share of famous strikes to his name, said ``It's very light, the way they are doing it is completely different from before. It seems like it's made of plastic."

The 2010 FIFA World Cup in South Africa then brought us the most controversial ball to date, the Jabulani.  The surface of the ball was constructed with 8 thermally bonded panels, each possessing a microtexture.  The criticism this time was especially widespread.  Brazilian keeper Julio Cesar said ``It's terrible, horrible. It's like one of those balls you buy in the supermarket'' while the English custodian Joe Hart described it doing ``anything but staying in my gloves''.  Outfield players were also critical.  The Dane Daniel Agger said         
``it makes us look like drunken sailors'' while even the Brazilian forward Robinho said ``For sure the guy who designed this ball never played football''.  Its general reception by most players was sufficiently negative to prompt both FIFA and Adidas to release official comments on the ball, along with a promise for further dialogue.  This was an unprecedented measure.
There were however some memorable goals scored with the ball, perhaps none more than Maicon's goal in the group stage against North Korea~\cite{MaiconGoal}, in which the ball is seen to swerve significantly over a relatively short distance.   

The recent 2014 FIFA World Cup has come and gone, and the official match ball, the ``Brazuca'', with its 6 thermally bonded panels has undoubtedly been put under heavy scrutiny by the players.  Given the extensive criticism received by previous world cup balls, perhaps one wouldn't be surprised to hear a raft of complaints levelled at the Brazuca.  However, while there was certainly controversy on the field, there was generally praise for the ball, with the international media largely focusing on on-field issues.  With the 2014 Brazuca seemingly a well-received ball, the fans could easily be left wondering why there were so many complaints at previous world cups.  Were the player's complaints valid with previous balls being, in some sense, ``horrible'', or is it simply an easy excuse to use the ball as a scapegoat for any failures, past or potential?
A discussion of the physics involved in the ball's flight will shed light on an answer to such questions.   

In this paper we examine the aerodynamics of the three most recent world cup balls, the Teamgeist, the Jabulani and the Brazuca, comparing their aerodynamic properties to the 32-panel balls players are accustomed to from various leagues around the world.  We begin in Section \ref{sect:FundamentalBallAerodynamics} by outlining the basic physics at work as a ball flies through the air.  We then introduce a new functional form in order to fit drag coefficient ($C_{D}$) data in Section \ref{sect:DragCoefficients}.  Using these fits and our understanding of the physics of the boundary layer we then discuss the implications for the lift coefficient ($C_{L}$) of the ball in section \ref{sect:LiftCoefficients}, making use of empirical lift coefficient data where available.  In doing so, we develop a method to estimate the behaviour of lift coefficients in the absence of readily available empirical data, enabling us to produce accurate flight trajectories.  The effect altitude and temperature has on flight paths is then quantified in Section \ref{sect:Altitude}, where we focus on the two most recent world cup balls.  Finally throughout Section \ref{sect:SimulationResults} we present visualisations comparing flight trajectories with a variety of different balls and various initial conditions.

\section{Fundamental Ball Aerodynamics}
\label{sect:FundamentalBallAerodynamics}
As a ball flies through the air it is subject to forces due to gravity and the air through which it is flying.  We follow the usual practice~\cite{CarreMJAsaiTAkatsukaTandHaakeSJ, JEGoff, Goffetal2017} of separating the force exerted by the air into drag and lift forces, $\textbf{F}_D$ and $\textbf{F}_L$ respectively, as shown in Fig. 1.

For a given ball the drag force is in a direction opposite to the ball's velocity and has magnitude~\cite{DeMestreN}

\begin{equation} \label{eq:Fd}
\textrm{F}_D = \frac{1}{2}\,C_D(Re, S\!{\kern 0.125em}p)\,{\rho}\,A\,v^{2},
\end{equation} 
while the lift force, or Magnus force, is in the direction $\vec{\omega} \times \vec{v}$ and has magnitude~\cite{DeMestreN} 

\begin{equation} \label{eq:FL}
\textrm{F}_L = \frac{1}{2}\,C_L(Re, S\!{\kern 0.125em}p)\,{\rho}\,A\,v^{2}.
\end{equation}
Here $\vec{\omega}$ is the ball's angular velocity, $v$ is the magnitude of the ball's velocity, $\vec{v}$, $\rho$ is the air density, and $A$ is the ball's cross sectional area.  $C_D$ and $C_L$ are the ball's drag and lift coefficients respectively. It's interesting to note that the lift force was first observed by Robins~\cite{Robins} in measuring the trajectories of spinning projectiles.  Reference~\cite{MagnusCylinder} provides a contemporary review.  

These coefficients are functions of Reynolds number \textit{Re}, and spin parameter \textit{Sp}, where

\begin{equation} \label{eq:Re}
Re :=  \frac{v\,D}{\nu_{k}},
\end{equation}
and 
\begin{equation} \label{eq:Sp}
S\!{\kern 0.125em}p :=  \frac{r\,\omega}{v}.
\end{equation}
%
\setlength{\unitlength}{4cm}
\begin{center}
\begin{figure}
\begin{center}
\begin{minipage}[c]{12.25cm}
\framebox{
\begin{picture}(3,2)
\thicklines
\put(1.5,1){\circle{1.5}}
\put(1.5,1){\vector(3,2){0.8}}
\put(1.5,1){\vector(-2,3){0.5}}
\put(1.5,1){\vector(-3,-2){0.8}}
\put(1.5,1){\vector(0,-1){0.8}}
\thinlines
\multiput(0.25,1)(0.2,0){13}
{\line(1,0){0.1}}
\qbezier(1.99,1.32)(2.12,1.22)(2.1,1)
\put(2.01,1.3){\vector(-1,1){0.01}}
\put(2.1,1.55){$\textbf{v}$}
\put(1.94,1.09){$\theta$}
\put(1.58,0.2){$\textbf{F}_G = m\textbf{g}$}
\put(0.55,0.6){$\textbf{F}_D$}
\put(1.1,1.7){$\textbf{F}_L$}
\end{picture}
}
\caption{A schematic diagram of the forces on a ball travelling with velocity $\textbf{v}$ at angle $\theta$ to the ground (magnitudes are of course not to scale).  Here the lift force is drawn for backspin, $\vec{\omega}$, with the spin axis orthogonal to the plane of the page and pointing out of the page.}\par
\end{minipage}
\end{center}
\end{figure}
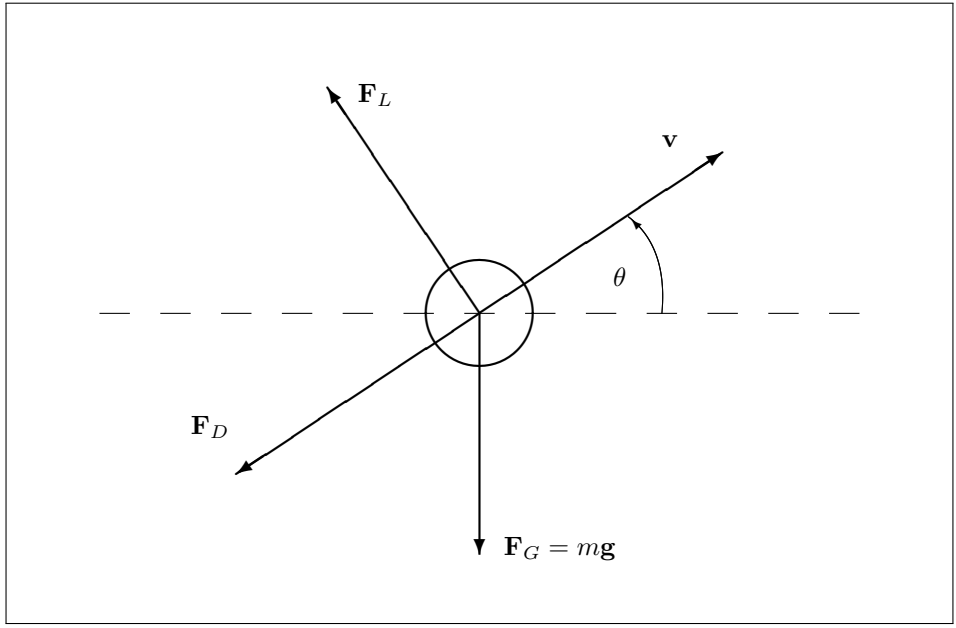
\end{center}

Here r is the ball's radius, $D = 2r$ its diameter and $\omega$ its angular speed, $|\vec{\omega}|$.  $\nu_{k}$ is the kinematic viscosity which is given by the ratio of viscosity to air density.  Crucially, $C_D$ and $C_L$ are also dependent on the ball's surface properties~\cite{MehtaRD, AchenbachEhighRe,AchenbachEroughsurf, MehtaRDandPallisJM} via the physics of the thin layer of air around the ball, called the boundary layer.  
%
%
\begin{figure}[t!!] 
  \begin{center} 
  \captionsetup[subfigure]{width=0.42\textwidth}
  \subfloat[A diagram showing the boundary layer separating early in the low speed and high drag laminar flow regime creating a large wake.]
  {\includegraphics[width=0.46\textwidth]{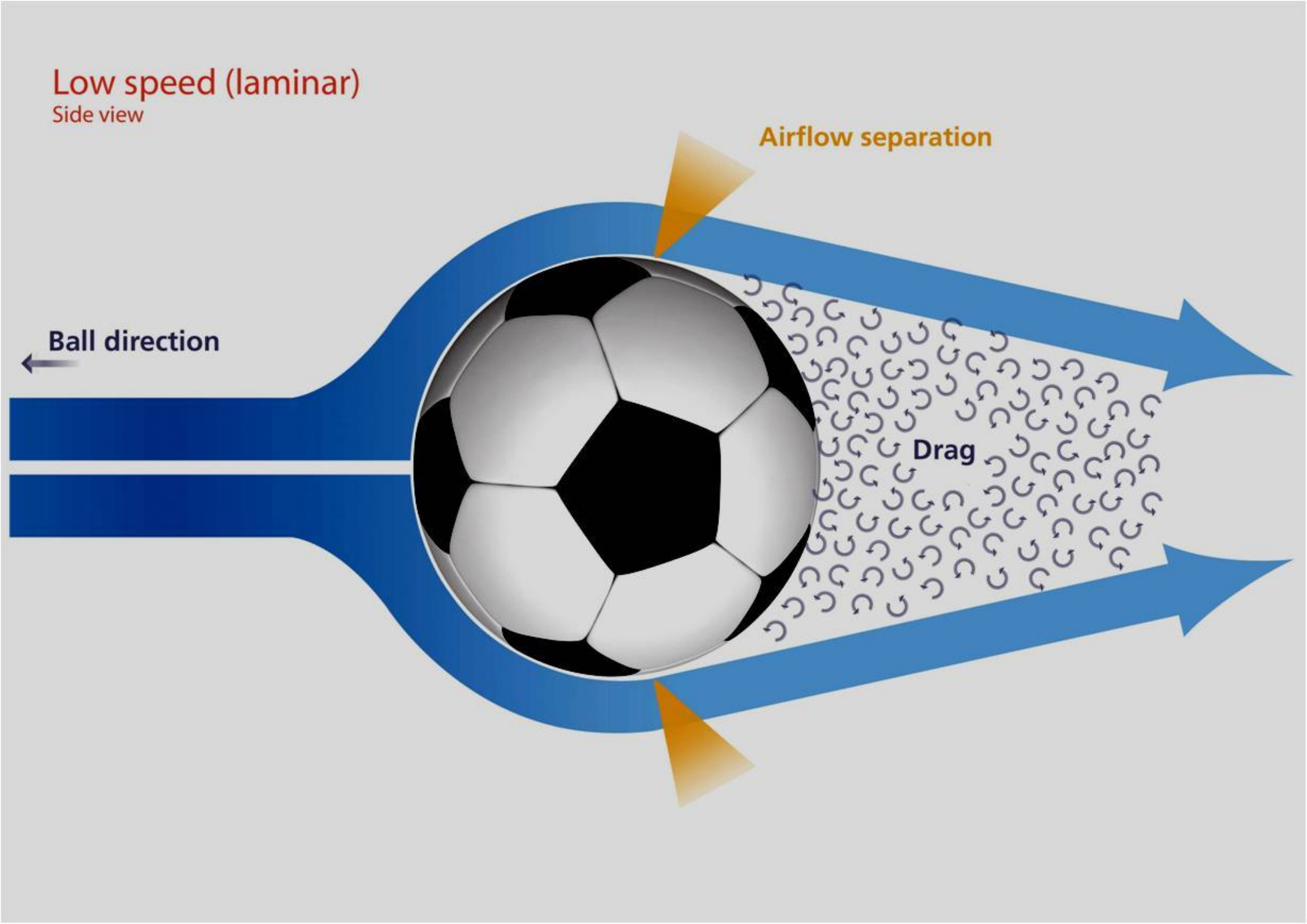}}
   \subfloat[A diagram showing the boundary layer separating late in the high speed and low drag turbulent flow regime creating a small wake.]
  {\includegraphics[width=0.46\textwidth]{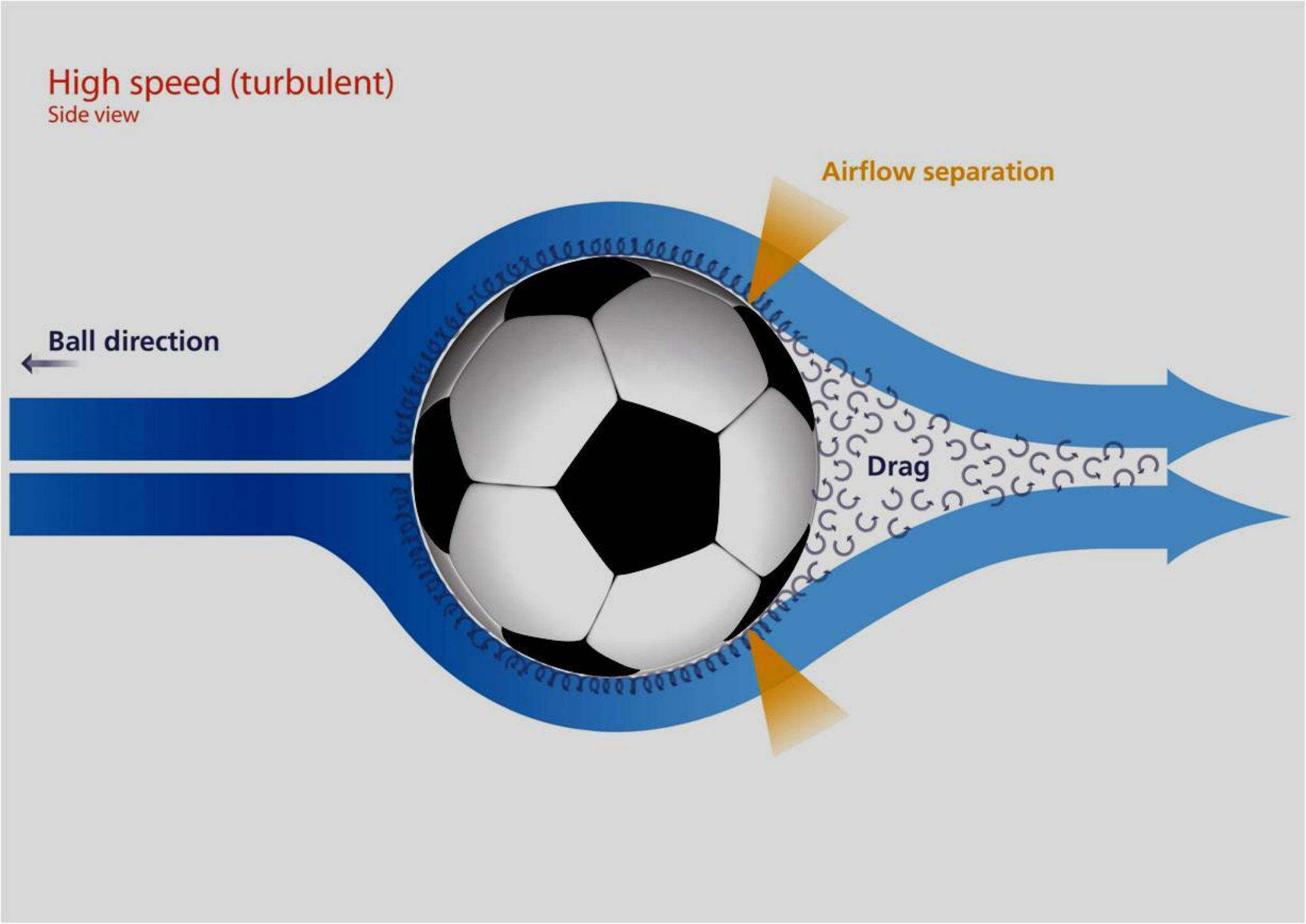}}
 \end{center}
\vspace{-0.5cm}
 \caption{(Colour online) Diagrams showing boundary layer separation for laminar and turbulent regimes. }
 \label{fig:BoundaryLayerComparison}
\end{figure}
%
%
At low speeds, in the laminar airflow regime, the boundary layer separates early creating a large wake with high drag (high $C_D$) as shown in part a) of Figure \ref{fig:BoundaryLayerComparison}.  On the other hand, at high speeds, in the turbulent airflow regime, the boundary layer separates late from the ball creating a small wake and hence low drag (small $C_D$) as seen in part b) of Figure \ref{fig:BoundaryLayerComparison}. 
Wind tunnel visualisations of this effect can be found in references~\cite{AsaiTSeoKKobayashiOandSakashitaR, ThorstenKetal}.  The speed at which this turbulent-to-laminar transition occurs, called the critical Reynolds number, is directly affected by the surface roughness of the ball.  It has been shown for spheres that high levels of surface roughness correspond to a relatively low critical Reynolds number~\cite{AchenbachEroughsurf}.  This surface roughness serves to trip turbulence, delaying the onset of the laminar regime as the ball slows.  However, the same roughness also serves to increase drag in the turbulent regime, by virtue of thickening the boundary layer.  The critical speed at which the turbulent-to-laminar transition occurs can have wide-reaching consequences for a ball's aerodynamic performance, introducing the possibility of surprising even the game's best goalkeepers.  This is discussed in more depth for each ball individually in Section \ref{sect:DragCoefficients}.

\begin{figure}[t] 
  \centering
  {\includegraphics[width=0.74\textwidth]{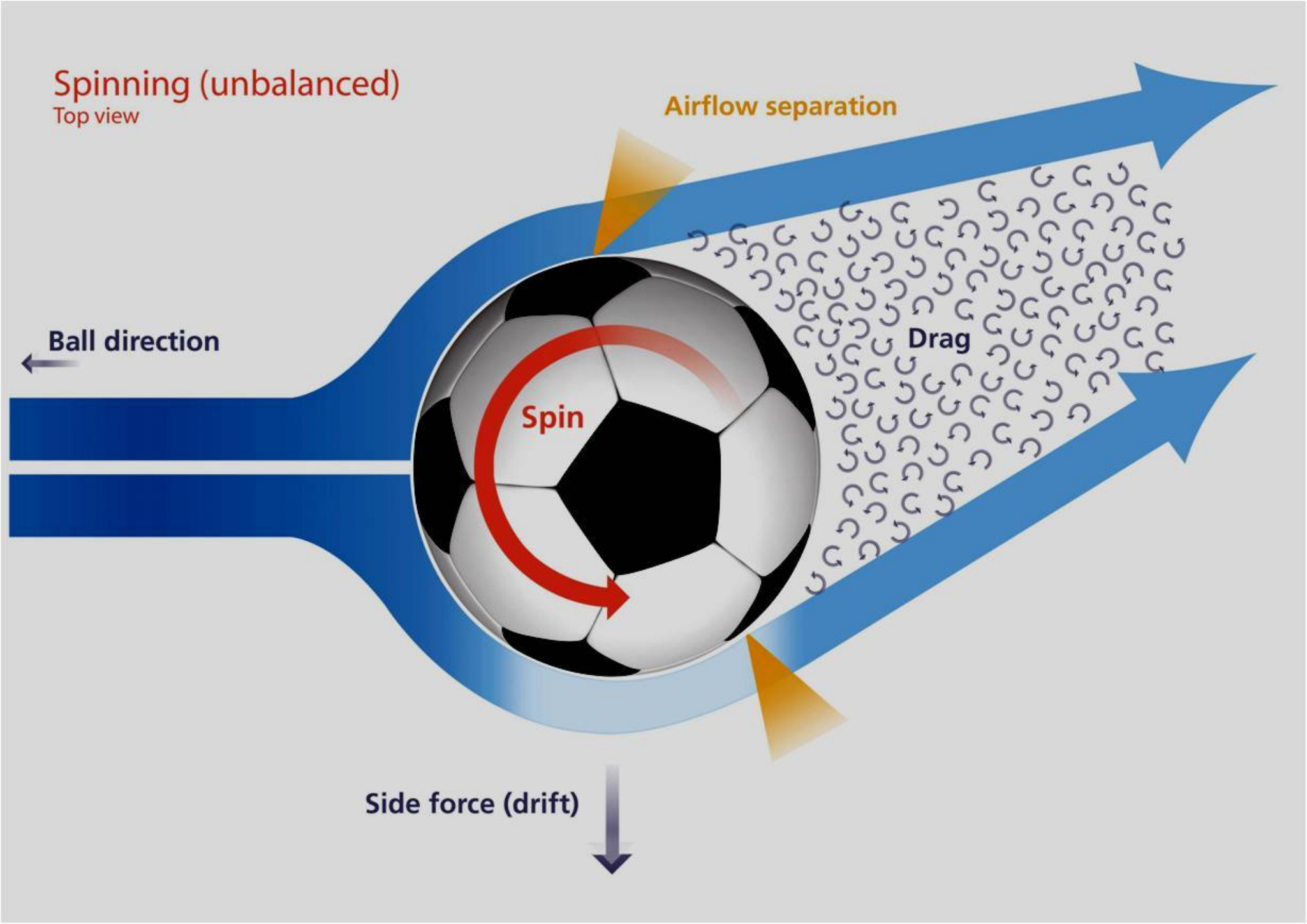}}
  \caption{(Colour online) A diagram showing the boundary layer properties around a rotating ball.  The lighter shading of the boundary layer represents the lower pressure of the Bernoulli effect.}
 \label{fig:SpinningBall}
\end{figure}

The turbulent-to-laminar transition is not the only difficulty goalkeepers have to contend with.  A common technique when shooting for goal is to impart spin on the ball in order to perform a swerving shot.  A spinning ball drags the boundary layer with it resulting in later boundary layer separation on one side of the ball than the other, creating an uneven wake as shown in Figure \ref{fig:SpinningBall}.  This uneven boundary layer separation unbalances the side forces on the ball contributing to the Magnus force, and hence to the swerve of the ball.  In the ball's reference frame the air in the boundary layer is travelling at $\backsim v - r\omega$ on one side and $\backsim v + r\omega$ on the other.  By the Bernoulli principle faster flowing air creates a region of lower pressure, meaning a spinning ball will create a pressure gradient pulling the ball further ``towards its nose''.  The pressure gradient and the uneven boundary-layer separation work together to produce a strong side force.  While back-spinning balls experience lift, side-spinning balls swerve.  In both cases it is common to refer to this Magnus force as a lift force.  

Furthermore, at low drag, the boundary layer is close to the ball being well intact, enhancing the aforementioned lift affects, whereas at high drag a thicker boundary layer that has been spoiled will result in a reduction.  Consequently, the ball's surface roughness is once again intimately linked to its aerodynamic performance.      

%
%
%
%
Total seam length and depth are currently believed to be the dominant measures of surface roughness for footballs~\cite{AsaiTandSeoK}.  Of the balls studied for this work the Jabulani has the lowest total seam length of $\backsim\textrm{203 cm}$ while the seam lengths of the Brazuca and Teamgeist are $\backsim\textrm{327 cm}$ and $\backsim\textrm{345 cm}$ respectively~\cite{ConversationArticle}.  Conventional 32-panel balls have a seam length of $\backsim\textrm{400 cm}$~\cite{AsaiTandSeoK, ConversationArticle}.  In addition to having the shortest seam length the Jabulani also has the most shallow seams of $\backsim\textrm{0.48 mm}$, while a conventional 32-panel ball's seams are $\backsim\textrm{1.08 mm}$ deep~\cite{ConversationArticle} and the Brazuca has the largest value of $\backsim\textrm{1.56 mm}$.  This makes the Jabulani the smoothest of all the balls, meaning it will have the highest critical Reynolds number.  Given the seam lengths and depths of the 2014 Brazuca and 32-panel balls above, we would anticipate them to have similar aerodynamic properties.

Currently, the sport's world governing body, FIFA, has relatively tight restrictions on multiple ball properties, such as circumference, mass and initial pressure of the ball~\cite{FIFA-rules}.  However, there are no such regulations on the surface roughness or material used for the ball's construction, enabling ball manufacturers to alter $C_{D}$ and $C_{L}$.  For a given ball $C_D$ and $C_L$ can be determined empirically, either through trajectory analysis~\cite{CarreMJAsaiTAkatsukaTandHaakeSJ, JEGoff, GoffJEandCarreMJ, BrayKandKerwinDG} or by analysing wind tunnel data~\cite{CarreMJGoodwillSRandHaakeSJ, AsaiTSeoKKobayashiOandSakashitaR}.   

\section{Drag Coefficients}
\label{sect:DragCoefficients}
To create accurate flight trajectories it is essential to quantify the drag and lift coefficients in a manner that captures all the physics governing their behaviour.
Motivated by wind tunnel data~\cite{HaakeSJGoodwillSRandCarreMJ} for the $C_D$ dependence on Reynolds number for various non-spinning balls, we propose a new term in the $C_D$ fitting function used by~\cite{JEGoff, GiordanoNJandNakanishiH}.  The new fitting function (at $S\!{\kern 0.125em}p = 0$), is given by
\begin{equation}\label{eq:CD}
\!\!\!\!\!\!C_D(v)\big|_{S\!{\kern 0.125em}p = 0} = \frac{a - b_{min}}{1 + \exp[(v - v_c)/v_s]} + b_{min} + \frac{v - v_{min}}{1 + \exp[-(v - v_{min})/v_s]}\frac{b_{max} - b_{min}}{v_{max} - v_{min}},
\end{equation}
where the third term is a new term constructed in order to reproduce the increase in drag at high Reynolds numbers~\cite{HaakeSJGoodwillSRandCarreMJ}.  Given that goal scoring shots are likely to spend a significant portion of their flight path in this region, the third term is a useful addition.  We note here that the complete functional form and all coefficients are chosen in order to characterise the degrees of freedom required to reproduce a close fit to available data such as the wind tunnel results seen in reference~\cite{HaakeSJGoodwillSRandCarreMJ} without over-fitting.  Therefore, each coefficient has a corresponding attribute it governs, which we now proceed to describe.

The variable $a$ governs the maximum value attained by $C_D$ for the given ball, $v_c$ governs the critical velocity at which the turbulent-to-laminar transition occurs and $v_s$ governs the slope and hence relative speed of the associated transition. The variables ($b_{min}$, $b_{max}$) and ($v_{min}$, $v_{max}$) control the minimum and maximum $C_D$ values and velocities respectively of the linear tail of the function in the turbulent regime.  

\begin{figure}[th!!] 
  \begin{center} 
  \captionsetup[subfigure]{width=0.42\textwidth}
  \subfloat[The fit to Brazuca drag coefficient data.  Data points were obtained from wind tunnel experiments~\cite{BrazucaPaper}.  Values have been averaged over two seam orientations.]
  {\includegraphics[width=0.46\textwidth]{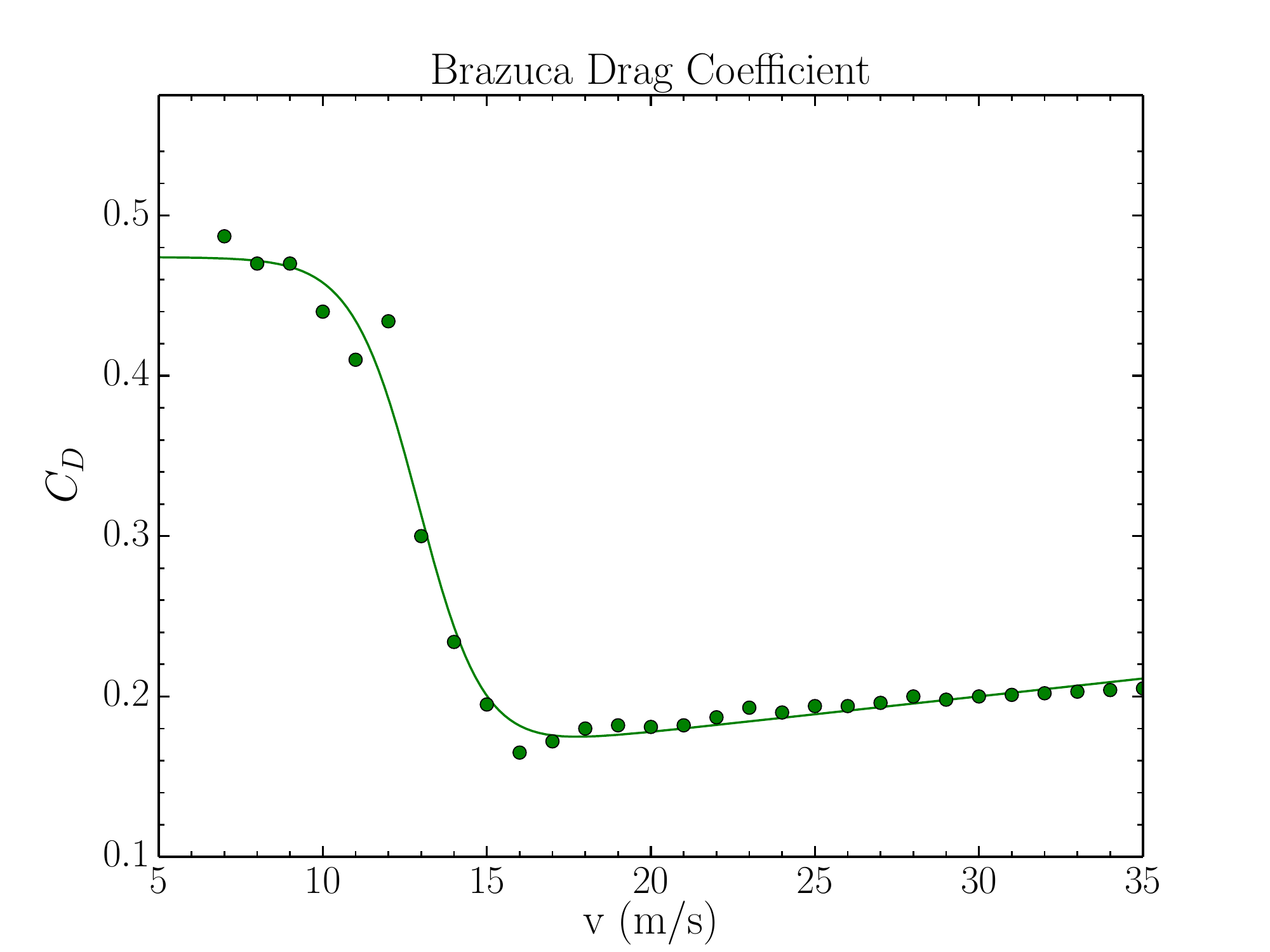}}
   \subfloat[The fit to Jabulani drag coefficient data.  Data points were obtained from wind tunnel experiments~\cite{BrazucaPaper}.  Values have been averaged over two seam orientations.]
  {\includegraphics[width=0.46\textwidth]{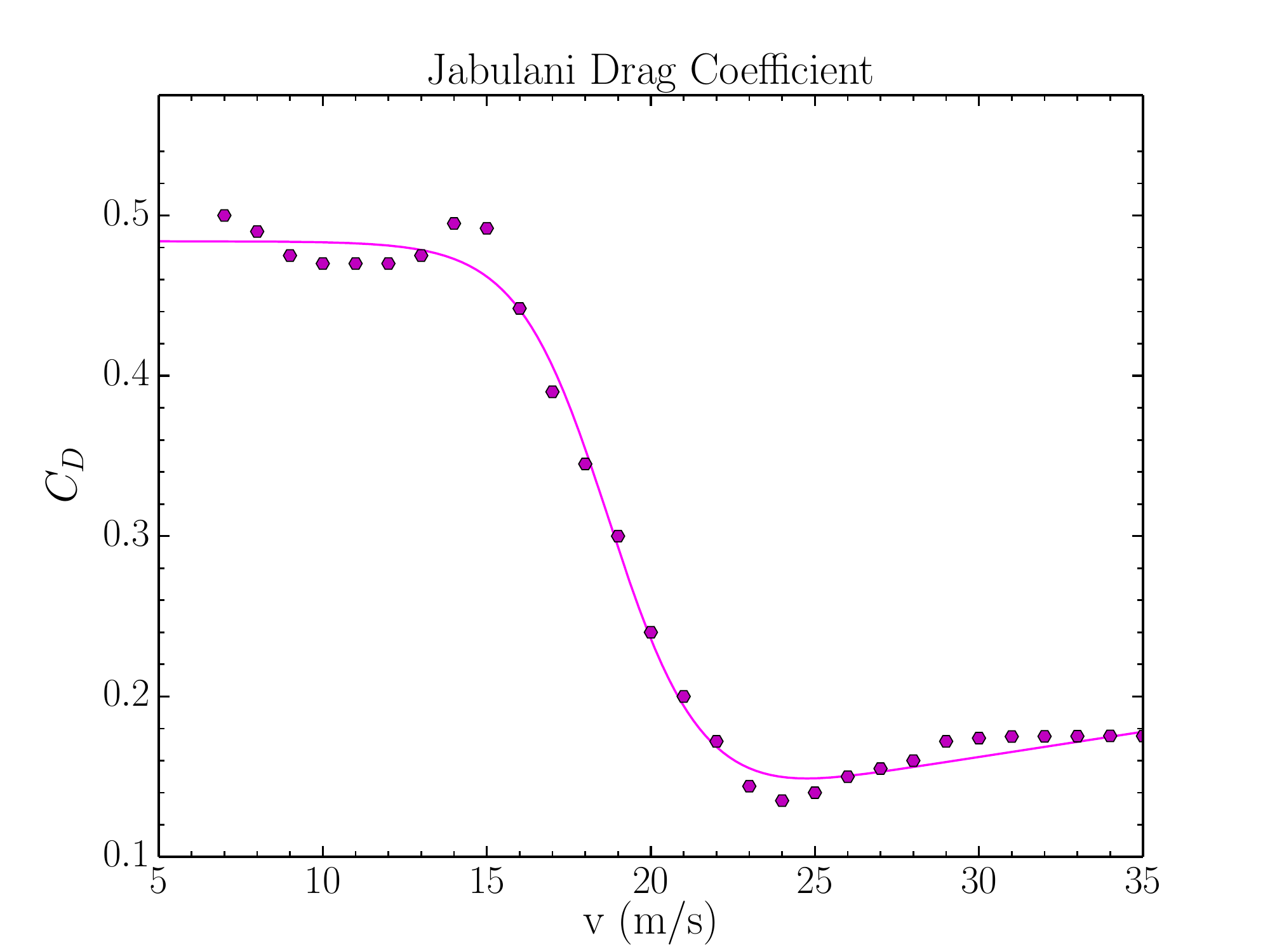}}\\  
  \subfloat[The fit to Teamgeist drag coefficient data.  Data points were obtained via trajectory analysis~\cite{JEGoff}.]
  {\includegraphics[width=0.46\textwidth]{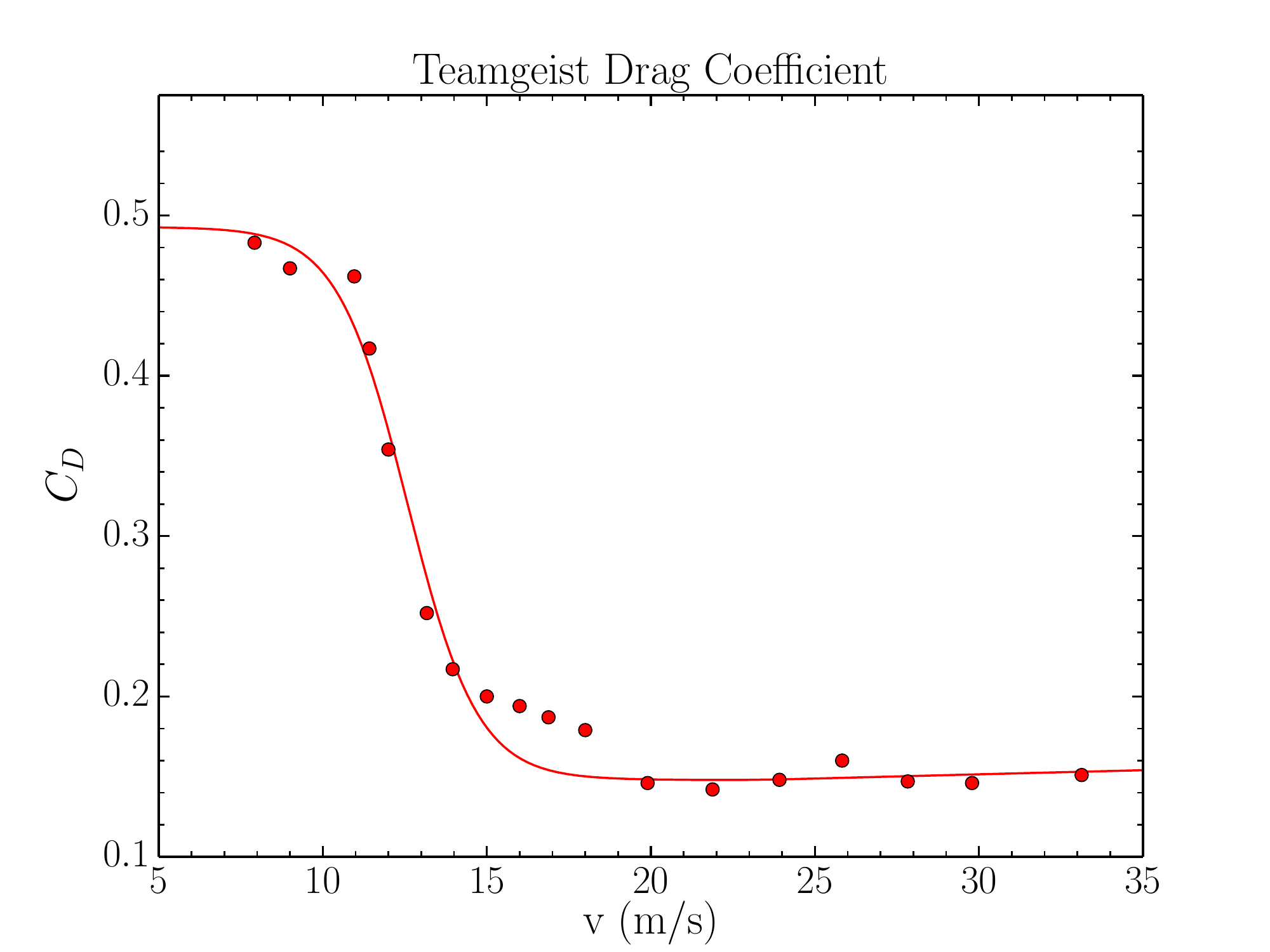}}
   \subfloat[The fit to Tango 12 drag coefficient data.  Data points were obtained from wind tunnel experiments~\cite{BrazucaPaper}.  Recall the Tango 12 is a 32-panel ball.]
  {\includegraphics[width=0.46\textwidth]{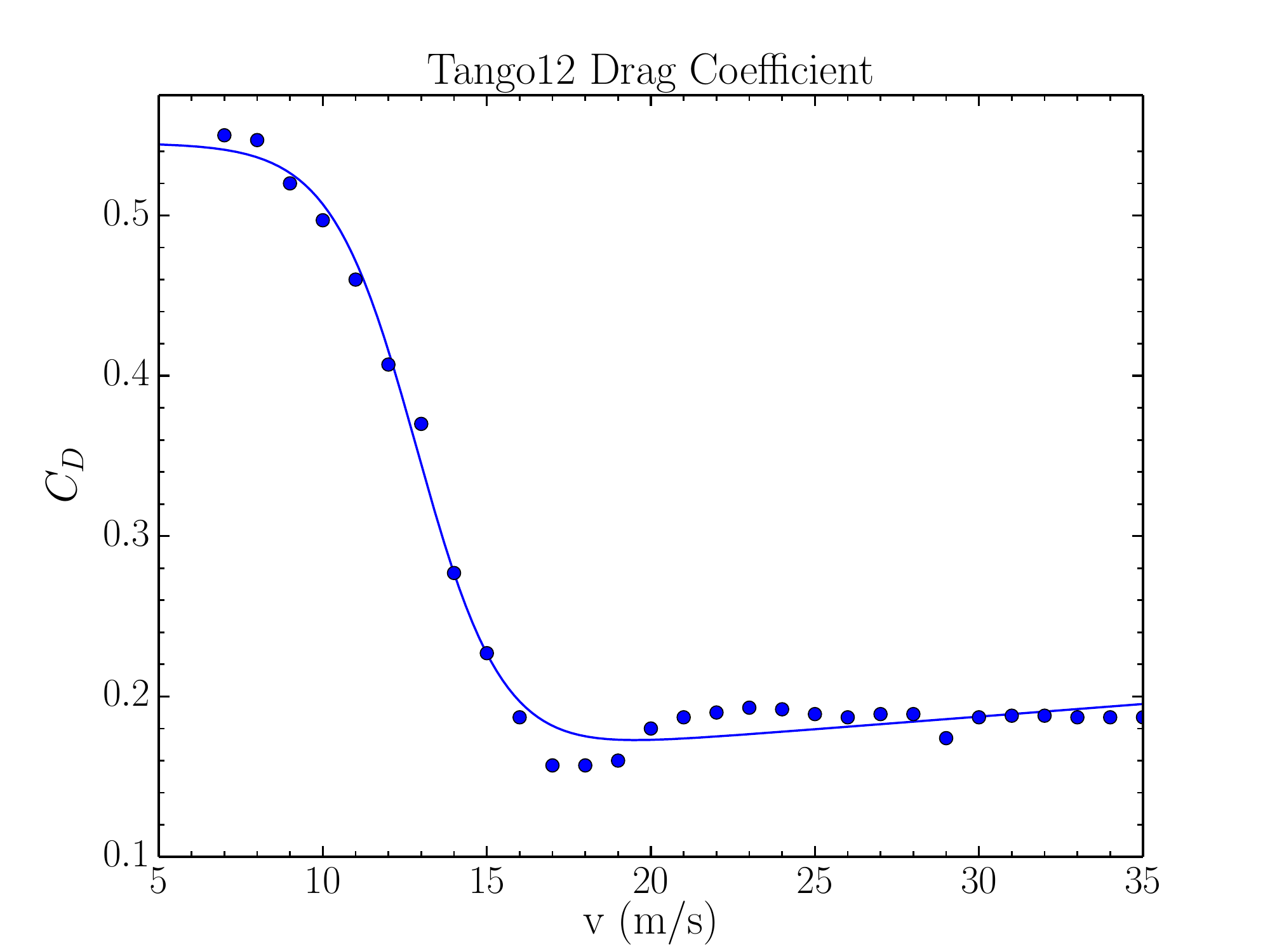}}
 \end{center}
\vspace{-0.5cm}
 \caption{(Colour online) Diagrams showing various drag coefficient data and their fits via Eq. \ref{eq:CD}.}
 \label{fig:CdFits}
\end{figure}

\begin{figure}[t] 
  \centering
  {\includegraphics[width=0.74\textwidth]{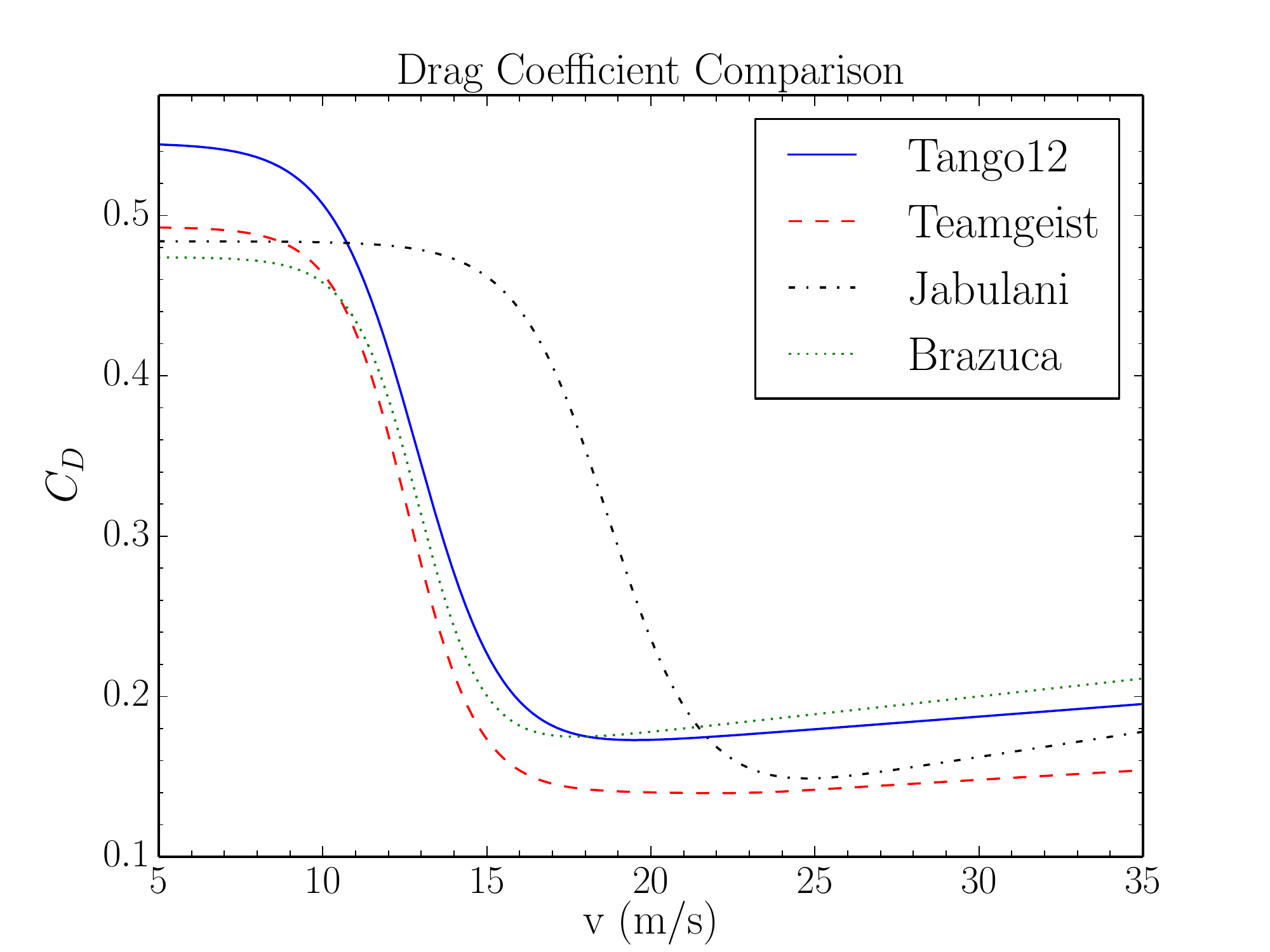}}
  \caption{(Colour online) A diagram comparing the fits for the drag coefficient of the three most recent FIFA world cup balls and the 32-panel Tango 12.  Note the extremely early turbulent to laminar transition for the Jabulani.}
 \label{fig:CdFitsCompare}
\end{figure}

Fits of Equation \ref{eq:CD} to available wind tunnel data are illustrated in Figure \ref{fig:CdFits}.
%
%
%
Figure \ref{fig:CdFitsCompare} confirms our expectations based on boundary layer physics and surface roughness as discussed in Section \ref{sect:FundamentalBallAerodynamics}.  As the Jabulani has the shortest total seam length and the shallowest seams, the turbulent-to-laminar transition occurs at the highest speed.  This effect is significant and easily observed in Figure \ref{fig:CdFits}.  Although a typical goal scoring shot would likely begin its trajectory at about 35 ms$^{-1}$, some shots could slow to as low as $\backsim 15$ ms$^{-1}$, meaning a portion of the trajectory is spent transitioning from turbulent to laminar boundary layer flows.  The Tango12 32-panel ball is next to undergo the transition to laminar flow followed closely by the Brazuca and Teamgeist.  This is in accord with our expectations based on surface roughness in Section \ref{sect:FundamentalBallAerodynamics}.  

During the turbulent-to-laminar transition, the airflow on one side of the ball may be slow enough to be laminar, while a seam or surface roughness may be tripping turbulence in the boundary layer on the opposite side.  The boundary layer on the turbulent side will separate later, giving rise to an uneven wake and hence a lift force.  As the ball is travelling at a speed in the critical region a small change in seam orientation may reverse the uneven wake with the ball now experiencing a lift force in the opposite direction.  It is exactly this type of erratic behaviour that makes a ball susceptible to knuckle-ball type effects that are well known from sports such as baseball~\cite{HongSetal, BaseballRef}. 
In this way the critical Reynolds number can be thought of as a characteristic speed near which the flight path of the ball becomes unpredictable.   

As shown in Figures \ref{fig:CdFits} and \ref{fig:CdFitsCompare}, this characteristic range of speeds varies from ball to ball.  For the Jabulani this unpredictable region lies approximately between $15 - 24$ ms$^{-1}$, while the the other balls have their corresponding region at approximately $10 - 17$ ms$^{-1}$.
Recall that while shots at goal can start their flight paths at speeds of up to $35$ ms$^{-1}$ just after the ball leaves the player's boot, the ball can slow to between $15 - 20$ ms$^{-1}$ by the time the ball arrives at the goaline and the keeper is called to make a save.  

This is precisely the reason for the much publicised complaints about the Jabulani.  Its unpredictable region where it was susceptible to knuckle-ball type effects happens to coincide with the typical speeds the ball was travelling at just before it arrived at the keeper or on the head of a striker.  As the ball can slow to $15 - 20$ ms$^{-1}$ during relevant match situations, all balls other than the Jabulani generally spend their flight path in the turbulent regime, making them in some sense predictable.  Furthermore, as the lift force is proportional to $v^2$, the force on the Jabulani when it starts to transition to a laminar flow is greater than that of the other balls.  For example, as the Jabulani begins to transition to a laminar flow it experiences a lift force approximately $({24}/{17})^2 \approx 2$ times greater than the lift force on the Brazuca when it begins its transition.  This will only serve to enhance the ball's erratic behaviour.

%
\begin{figure}[t] 
  \centering
  {\includegraphics[width=0.74\textwidth]{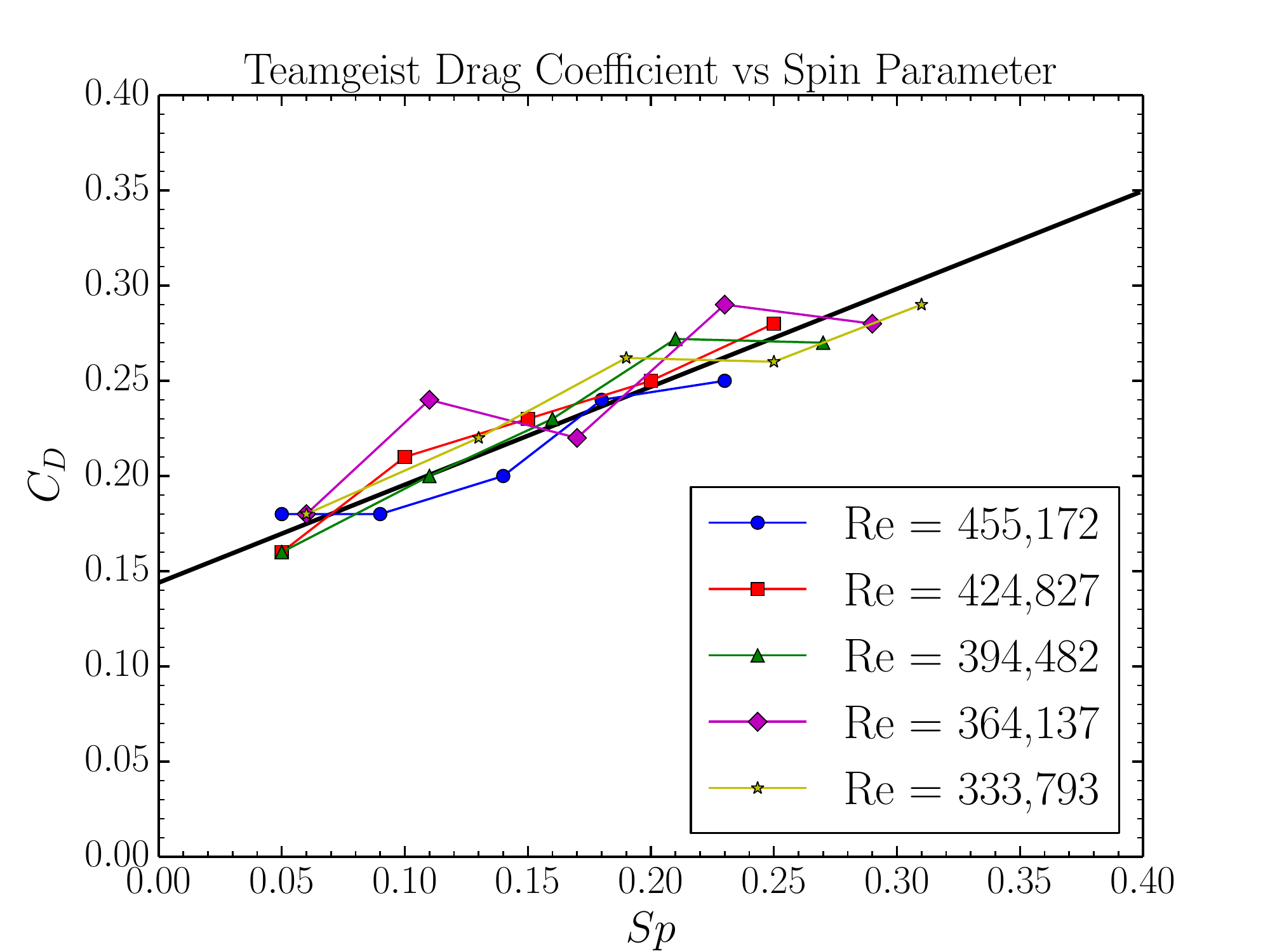}}
  \caption{(Colour online) A diagram quantifying the dependence of $C_{D}$ on $S\!{\kern 0.125em}p$.  Wind tunnel data has been gleaned from reference~\cite{AsaiTSeoKKobayashiOandSakashitaR}.  The solid black line shows the fit.}
 \label{fig:CdvsSpTeamgesit}
\end{figure}

%
%
%
\begin{table}[tb!!]
\centering
\begin{tabular}{c|c|c|c|c|c|c|c}
 \hline
\noalign{\vspace{3pt}}
    Ball & $a$ & $v_c$ & $v_s$ & $b_{min}$ & $b_{max}$ & $v_{min}$ & $v_{max}$  \\
\noalign{\vspace{3pt}}
    \hline 
    \hline
\noalign{\vspace{3pt}}
    Tango12 & 0.5452 & 12.86 & 1.304 & 0.1657 & 0.1953 & 16.22 & 35.00\\\hline
    Teamgeist & 0.4927 & 12.58 & 1.071 & 0.1440 & 0.1540 & 23.17 & 35.00\\\hline 
    Jabulani & 0.4839 & 18.69 & 1.377 & 0.1413 & 0.1780 & 23.29 & 35.00\\\hline 
    Brazuca & 0.4740 & 12.92 & 1.000 & 0.1657 & 0.2112 & 14.61 & 35.00\\\hline

\end{tabular}
\caption{A table showing the values of the fit parameters to the fitting function in Eq. \ref{eq:CD}.  These fits are shown in Figures \ref{fig:CdFits} and \ref{fig:CdFitsCompare}.}
\label{table:fit}
\end{table}
%
%
%
%
%
As we have seen in Section \ref{sect:FundamentalBallAerodynamics}, $C_D$ is not only a function of Reynolds number but also spin parameter.  While it is not the dominant contribution this dependence on spin parameter plays an important role in the determination of $C_D$.  The available empirical data documenting this dependence is not as abundant as studies on variations of $C_{D}$ with $Re$. However, there is wind tunnel data for the Teamgeist at a variety of different Reynolds numbers~\cite{AsaiTSeoKKobayashiOandSakashitaR} which we show in Figure \ref{fig:CdvsSpTeamgesit}.
%
%
%
%
The solid black line shows the fit to $a + bx$ with $a=0.144$ and $b=0.514 \textrm{ rad}^{-1}$.  Note that all the Reynolds numbers tested are in the turbulent regime and therefore show little flow-rate dependence.   
Furthermore, at zero spin parameter, $C_D$ is given by its value in the non-spinning case, that is, the value read from Figure \ref{fig:CdFitsCompare}.  A typical spin parameter for a swerving free-kick is approximately $0.2$ rad~\cite{GoffJEandCarreMJ}, while spin parameters above $0.4$ rad for flight trajectories are rare.  We note that for spinning balls, increased surface roughness is anticipated to more easily spoil the boundary layer, and therefore we expect rougher balls to have higher $b$ values.  For the purposes of our simulations in Section \ref{sect:SimulationResults}, and in the absence of experimental data, we restrict this enhancement in $b$ to $5$\% and linearly interpolate between the ball with the lowest drag coefficient (the Teamgeist), and the ball with the highest drag coefficient (the Brazuca).  We perform this interpolation at the typical initial free kick speed of $30$ ms$^{-1}$.  That is, 
\begin{equation}\label{Eq:bcalculation}
b = b_{teamgeist}\left(1 + 0.05\,\frac{C_{D} - C_{D}^{min}}{C_{D}^{max} - C_{D}^{min}}\right).
\end{equation}
Here, $C_{D}^{min}$ is smallest measured drag coefficient at $30$ ms$^{-1}$, (that of the Teamgeist), $C_{D}^{max}$ is the largest measured drag coefficient at $30$ ms$^{-1}$, (that of the Brazuca), $C_{D}$ is the drag coefficient of the ball in question measured at $30$ ms$^{-1}$, while $b$ and $b_{teamgeist}$ are the $b$ values of the ball in question and the Teamgeist respectively.  Consequently, the final value of $C_D$, which is calculated with the $b$ value from Equation \ref{Eq:bcalculation}, is then given by
\begin{equation}
C_D(Re, S\!{\kern 0.125em}p) = C_D\big|_{S\!{\kern 0.125em}p = 0} + b\,S\!{\kern 0.125em}p,
\end{equation}
where $C_D\big|_{S\!{\kern 0.125em}p = 0}$ is set by reading the appropriate value read from Figure \ref{fig:CdFitsCompare}.

%
%
%
\section{Lift Coefficients}
\label{sect:LiftCoefficients}
While drag coefficient data for various balls has been well documented, there is not as much empirical data available for lift coefficients.  Nevertheless, an understanding of the physics of the boundary layer enables the relevant lift coefficient information to be related to the drag coefficients.  In the turbulent regime the low drag corresponds to a small wake, meaning the boundary layer is intact.  The boundary layer in this regime is therefore close to the ball enhancing the effect of uneven boundary layer separation and pressure gradients that gives rise to lift.  Balls with lower drag coefficients will have higher lift coefficients. 

Conversely, as the ball undergoes transition to a high drag laminar flow the boundary layer is spoiled, separating much earlier creating a larger wake.  Given the boundary layer is now away from the ball, any uneven boundary layer separation or pressure gradient effects are diminished.  A reduction in the ball's lift coefficient occurs as it slows.

\begin{figure}[t] 
  \centering
  {\includegraphics[width=0.74\textwidth]{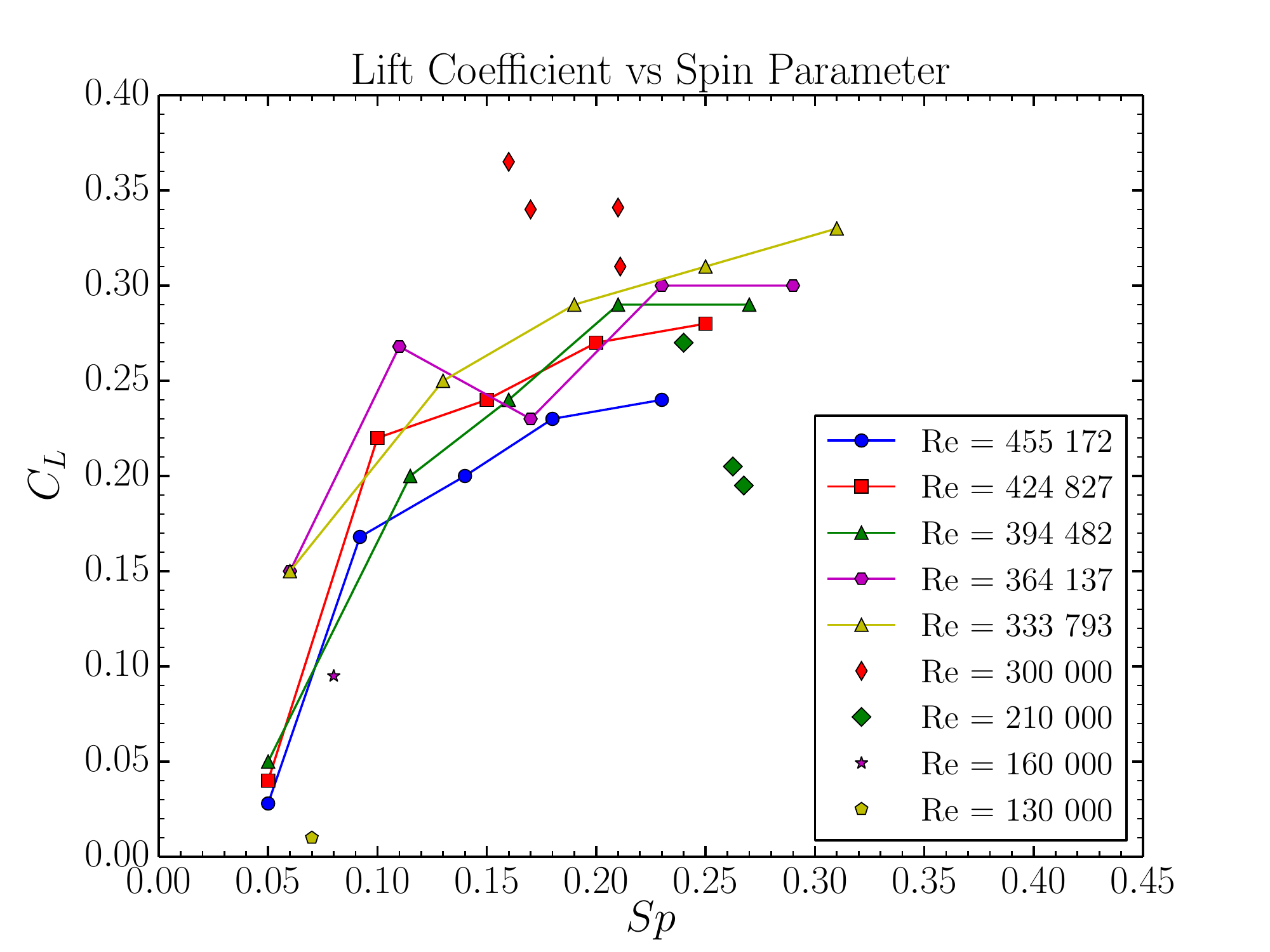}}
  \caption{(Colour online) A diagram showing the available lift coefficient, $C_{L}$, data for the Teamgeist.  The data points joined by a solid line are wind tunnel measurements~\cite{AsaiTSeoKKobayashiOandSakashitaR}.  Data points not joined by a line are sourced from a trajectory analysis~\cite{JEGoff}.}
 \label{fig:CLvsSpTeamgeist}
\end{figure}

Available data supports this understanding.  A compilation of relevant data from wind tunnel measurements and a trajectory analysis is shown in Figure \ref{fig:CLvsSpTeamgeist} for the Teamgeist ball.  Note that the turbulent-to-laminar transition for the Teamgeist occurs approximately between Reynolds numbers of $1.3 \times 10^{5}$ and $2.3 \times 10^{5}$, meaning all the wind tunnel data corresponding to the data points joined by the solid lines are in the turbulent regime.  The data points from the trajectory analysis are at a Reynolds number of $3 \times 10^{5}$.  We commence our discussion of Figure \ref{fig:CLvsSpTeamgeist} by turning our attention to the wind tunnel data.

There is a trend, particularly at higher values of the spin parameter, where higher Reynolds numbers generally correspond to lower drag.  Utilizing the boundary layer discussion of this section, this is consistent with the rising tails in the drag coefficient curves of Figure \ref{fig:CdFits}.  The Tango 12, Jabulani and particularly the Brazuca all display increasing drag with increasing Reynolds number within the turbulent regime.  
The effect is most pronounced for the Brazuca and the Jabulani which are the balls with the most surface roughness, whereas the glassy smooth panels of the Teamgeist better preserve the minimum drag observed.  Smooth panels preserve the boundary layer at high flow speeds whereas panel roughness acts to thicken and spoil the boundary layer~\cite{AchenbachEroughsurf, HaakeSJGoodwillSRandCarreMJ}.  Likewise, compromising the boundary layer compromises the lift.

As the Reynolds number is further decreased to $3 \times 10^{5}$ we observe an enhancement to the lift coefficient as expected.  Subsequent reductions to the Reynolds number correspond to speeds during which the turbulent to laminar transition is occurring, meaning the boundary layer is beginning to separate earlier.  We anticipate lower lift coefficients for the ``in transition'' Reynolds numbers of $2.1 \times 10^{5}$ and $1.6 \times 10^{5}$.  The final data point at a Reynolds number of $1.3 \times 10^{5}$ is now in the laminar regime as we consequently observe the lift coefficient dropping to near zero.

Turning our discussion toward the spread of the data, we note that the variation particularly between the trajectory analysis data points is quite considerable.  In order to obtain consistent results from a trajectory analysis, one must orientate the ball identically between successive trials.  This can be particularly difficult in practice leading to the observation of successive balls exiting the launcher with slightly different seam orientations~\cite{JEGoff}.  As previously discussed, a differing seam orientation can alter the manner in which the boundary layer separates from the ball thus altering the lift force.  The producers of these data points suggest overcoming this issue in future experiments by devising a method to visualise boundary layer flow during the launch.  A discussion of other minor contributions to this variation can be found in reference~\cite{JEGoff}.

We now draw on this understanding of the connection between drag and lift coefficients to model the anticipated flight trajectory of balls.  In order to model the behaviour of the lift coefficients of various balls, we begin by considering a power law of the form 
\begin{equation}
C_{L}^{fit}(S\!{\kern 0.125em}p) = \alpha\,{S\!{\kern 0.125em}p}{\kern 0.05em}^{\beta}.
\end{equation}
This choice is motivated by the data in Figure \ref{fig:CLvsSpTeamgeist}.  We choose a reference Reynolds number of 333 793 at which we perform our fit, since this is the smallest Reynolds number we have data for that remains in the turbulent regime, and consequently corresponds to the lowest $C_{D}$ and largest $C_{L}$.  As noted previously, we expect the low $C_{D}$ values associated with an intact boundary layer to correspond to high $C_{L}$ values and visa versa.  Consequently, we aim to perform a linear interpolation up to the maximum drag coefficient, at which point the boundary layer will be completely blown away giving rise to a lift coefficient of zero.  That is, we aim to encapsulate the behaviour of $C_{L}$ on $S\!{\kern 0.125em}p$ by fitting
\begin{equation}\label{eq:CLPrediction}
C_L(Re, S\!{\kern 0.125em}p) = C_{L}^{fit}(S\!{\kern 0.125em}p)\left(\frac{C_{D}|_{Re = 0} - \min(C_{D}, C_{D}|_{Re = 0})}{C_{D}|_{Re = 0} - C_{D}^{ref}}\right).
\end{equation} 
Here, $C_{L}^{fit}$ encapsulates the power behaviour of the $C_{L}$ dependence on $S\!{\kern 0.125em}p$ displayed in Figure \ref{fig:CLvsSpTeamgeist}, $C_{D}|_{Re = 0}$ is the drag coefficient at $Re = 0$ of the the ball whose data we are using to interpolate.  In our case we use the Teamgeist.  $C_{D}^{ref}$ is the drag coefficient evaluated at the reference spin parameter, (which we discuss imminently) and $C_{D}$ is the drag coefficient of the ball we are considering at $(Re, S\!{\kern 0.125em}p)$. 

We note here that the drag coefficients with which we are scaling $C_{L}^{fit}$ themselves depend on $S\!{\kern 0.125em}p$ via the final term of Equation \ref{eq:CD}.  Drawing from the data at the reference Reynolds number of 333 793, we aim to achieve $C_{L} = 0.33$ for $S\!{\kern 0.125em}p = 0.31$, and enforce the fit to pass through $(S\!{\kern 0.125em}p, C_{L}) = (0.19, 0.29)$, which we set to be our reference spin parameter.  Consequently, we discover $(\alpha, \beta) = (1.15, 0.83)$ not only encapsulates the trend of reduced lift at higher flow rates in the turbulent regime, but also produces predictions for the point at $S\!{\kern 0.125em}p = 0.06$ well (providing 0.14 compared with the empirical value of 0.15).  We therefore use these values of $\alpha$ and $\beta$ in our simulations.  

It is evident that the empirical data contains significant uncertainties, particularly at low values of $S\!{\kern 0.125em}p$, and future studies quantifying these uncertainties and/or increasing statistics would be useful in producing more accurate modelling of the data.
%
%
%
%
\section{Altitude}
\label{sect:Altitude}
In recent times the maximum altitude at which official matches can be held has been a topic of heated discussion.  This is particularly relevant on the south American continent where a significant portion of stadia are located at high elevations.  The 2007 Copa Libertadores, a South American international club competition, pitted Brazilian side Flamengo against Real Potos\'i of Bolivia, whose home stadium is 3,960 m above sea level.  Following a petition by the Brazilian Football Confederation outlining that players were making use of bottled oxygen at such venues, FIFA imposed a ban on international matches above 2,500 m in May 2007 meaning Bolivia, Colombia and Ecuador could no longer host World Cup qualifiers in their capital cities.  This limit was later lifted to 3,000 m meaning only Bolivia's La Paz was effected.  Ultimately however, in May 2008 the ban was suspended by FIFA after an official complaint from CONMEBOL, the governing body of south American football, which was supported by all member nations except Brazil.

While FIFA's concerns were primarily about the health of the players and consequently the integrity of the competition, altitude also has an influence on the ball's trajectory via changing air density.  As shown in Equations \ref{eq:Fd} and \ref{eq:FL}, the forces on the ball are directly proportional to air density which decreases with increasing altitude.

Air density, $\rho$, can be estimated with the ideal gas law~\cite{IdealGasLaw}
\begin{equation}
\rho = \frac{p\,M}{R\,T},
\end{equation}
where $M$ is the molar mass, $R$ is the ideal universal gas constant, $T$ the temperature and $p$ is the pressure which can be expressed as a function of elevation $h$ as
\begin{equation}
p = p_{0}\,\Bigg[1 - \frac{L\,h}{T_{0}}\Bigg]^{\frac{g\,M}{R\,L}}.
\end{equation}
Here $p_{0}$ is atmospheric pressure at sea level, $T_{0}$ is the temperature at sea level, $g$ is the gravitational acceleration and $L$ is the temperature lapse rate, defined to be the rate at which atmospheric temperature decreases with increasing altitude.  Naturally, $L$ will vary as a result of atmospheric temperature variations.  We note here that the ideal gas law is accepted as a valid approximation for air at Standard Temperature and Pressure (STP) of 0$^{\circ}$C and 101.3\,kPa.  At typical South American altitudes the pressure is lower and the temperature is higher than STP, and hence the air more closely approximates an ideal gas.

\begin{figure}[t] 
  \begin{center} 
  \captionsetup[subfigure]{width=0.42\textwidth}
  \subfloat[A plot of air density against elevation.  The temperatures of 12$^{\circ}C$ and 27$^{\circ}C$ are the average minimum and maximum temperatures in Bras\;ilia during the tournament months of June and July.  The lines at 3,600 m are shown by way of comparison with Bolivia's capital La Paz.]
  {\includegraphics[width=0.46\textwidth]{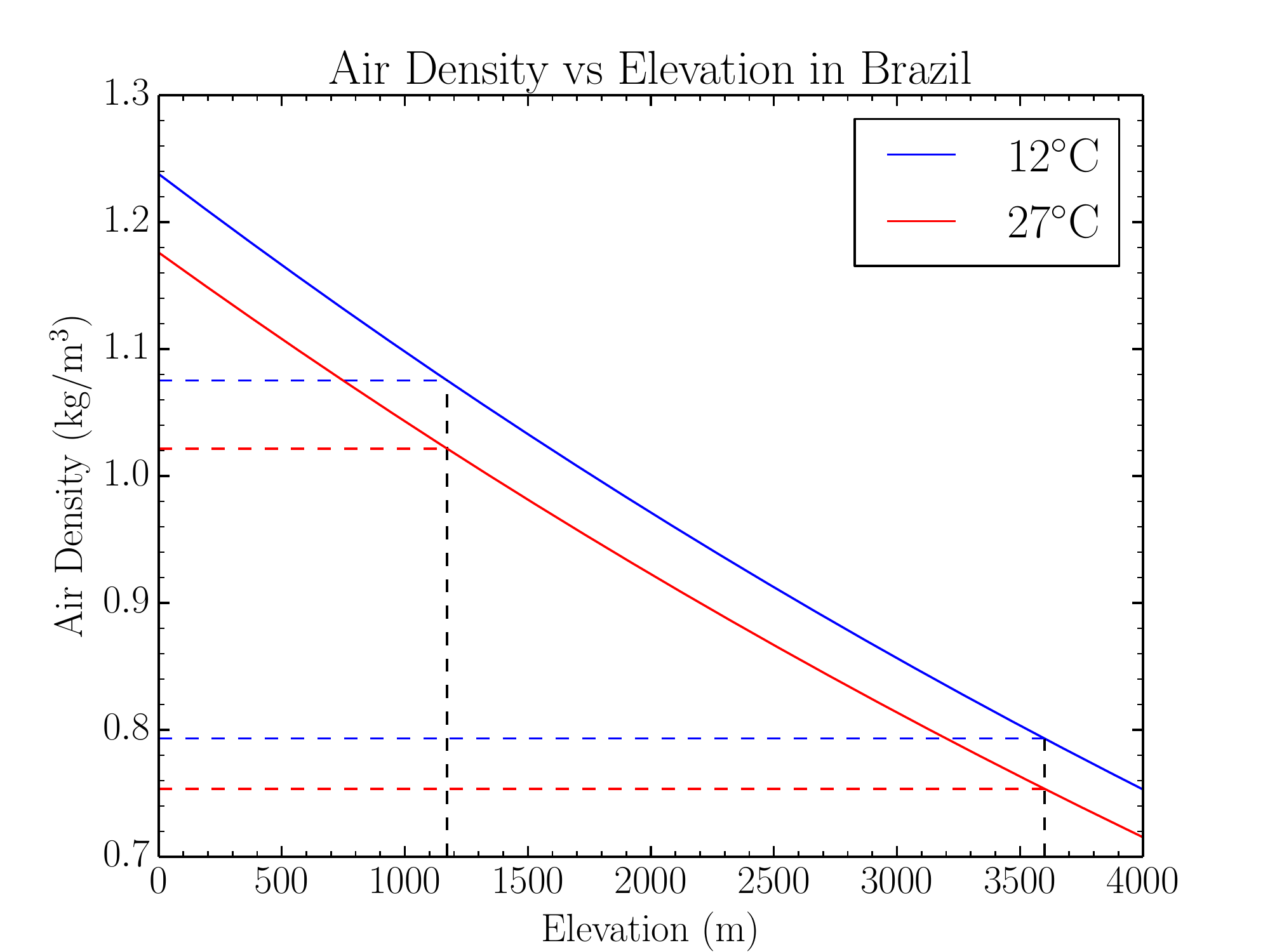}}
   \subfloat[A plot of air density against elevation.  The temperatures of 3$^{\circ}C$ and 18$^{\circ}C$ are the average minimum and maximum temperatures in Johannesburg during the tournament months of June and July.  The lines at 3,600 m are shown by way of comparison with Bolivia's capital La Paz.]
  {\includegraphics[width=0.46\textwidth]{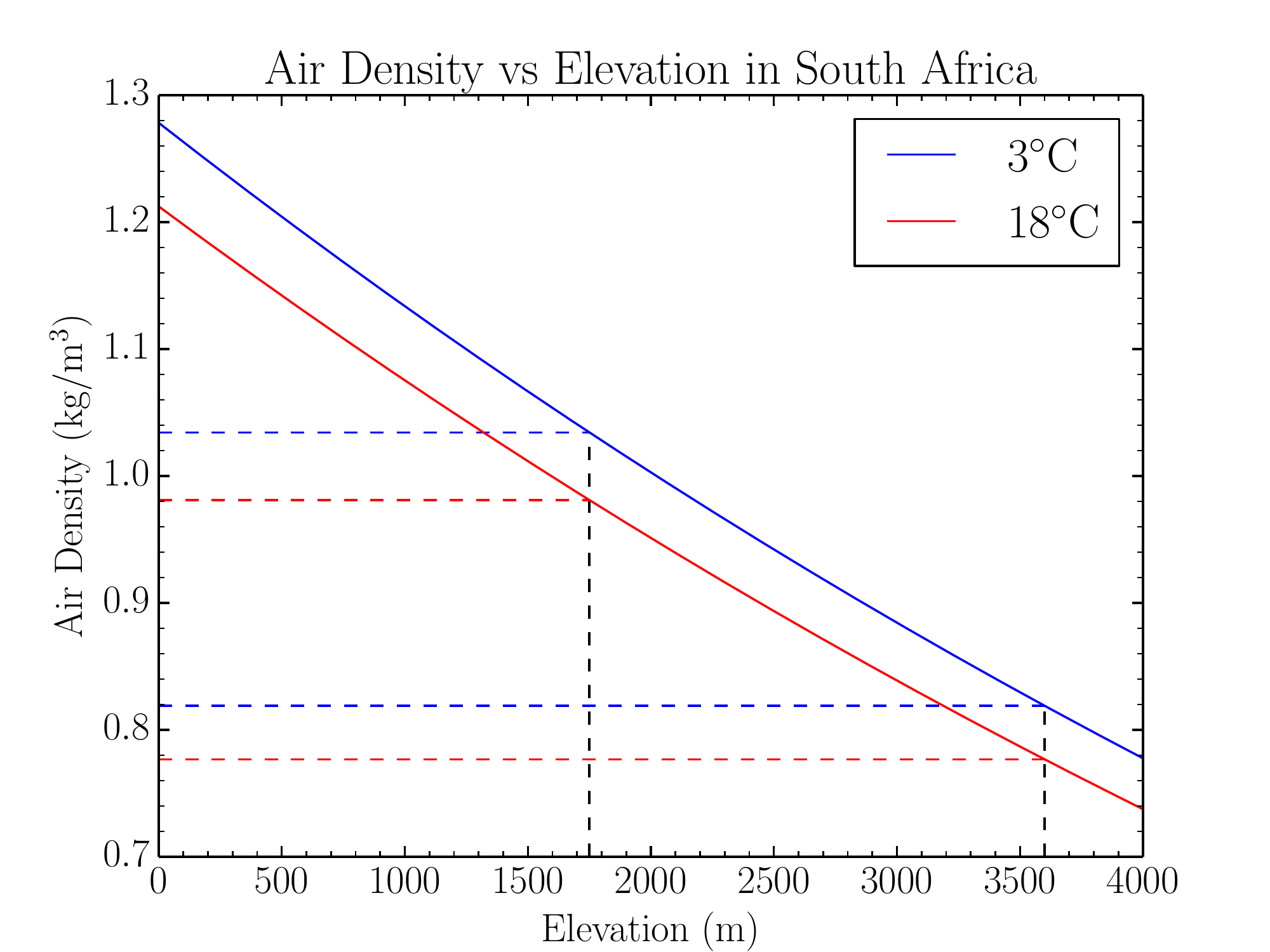}}
 \end{center}
\vspace{-0.5cm}
 \caption{(Colour online) Plots of Air Density against altitude.}
 \label{fig:AirDensityvsAltitude}
\end{figure}

In the 2014 World Cup in Brazil, the Est\'adio Nacional in Bras\'ilia was the highest altitude stadium at almost 1,200m above sea level.  Matches were also held in coastal cities and were consequently at sea level.  Figure \ref{fig:AirDensityvsAltitude} a) shows the variation in air density with altitude at 12$^{\circ}$C and 27$^{\circ}$C which are the average minimum and maximum temperature during the tournament months of June and July in Bras\'ilia \cite{WeatherWebsite}.  The lines at 3,600 m show the air densities at the same temperatures in the Bolivian Capital of La Paz for comparison.

This variation is smaller than it was during the 2010 tournament.  South Africa held matches in coastal cities near sea level, while games were also played in Johannesburg at an elevation of approximately 1,750 m.  Figure \ref{fig:AirDensityvsAltitude} b) shows the variation in air density with altitude at 3$^{\circ}$C and 18$^{\circ}$C which are the average minimum and maximum temperatures during June and July in Johannesburg~\cite{WeatherWebsite}.  Once again the lines at 3,600 m are for comparison with La Paz.
%
%
The effect on air density is directly proportional to the drag and lift forces and the amplitude is large with up to 23\% effects observed in Figure \ref{fig:AirDensityvsAltitude} over the temperature and altitude ranges encountered.

%
%
\section{Simulation Details}
\label{sect:SimulationDetails}
Prior to presenting simulation results, we here briefly outline the method by which we solve the relevant differential equations.  Using Newton's second law and the expressions for drag and lift forces in Section \ref{sect:FundamentalBallAerodynamics}, we sum the forces on the ball to obtain the differential equation governing the ball's flight which is given by
\begin{equation}
\frac{d^2{\vec{r}}}{dt^2} = -g\hat{\jmath} - \frac{\rho{A}v^2}{2m}\bigg[C_{D}(Re,S\!{\kern 0.125em}p)\,\hat{v} - C_{L}(Re,S\!{\kern 0.125em}p)\,(\hat{s}\times\hat{v})\bigg].
\end{equation}  
Here $\hat\jmath$ is the unit vector in the $y$ direction, $\hat{v}$ is the unit vector in the direction of the ball's motion, and $\hat{s}$ the unit vector on the ball's rotation axis.  It is worth noting that the aerodynamic response to surface roughness is encoded within the $C_D$ and $C_L$ coefficients discussed in Sections \ref{sect:FundamentalBallAerodynamics}, \ref{sect:DragCoefficients} and \ref{sect:LiftCoefficients}.  We solve this via the implementation of the iterative 5$^{th}$ order Cash-Karp Runge-Kutta algorithm, embedded with a 4$^{th}$ order approximation, enabling error estimates to be performed.

The general form for a 5$^{th}$ order Runge-Kutta is
\begin{equation}
y_{n+1} = y_{n} + c_{1}k_{1} + c_{2}k_{2} + c_{3}k_{3} + c_{4}k_{4} + c_{5}k_{5} + c_{6}k_{6} + \mathcal{O}(h^6),
\end{equation}
where the value of $k_{i}$ are given by
\begin{align}
k_{1} &= hy^{\prime}(x_{n},y_{n})\nonumber\\
k_{2} &= hy^{\prime}(x_{n} + a_{2}h,y{n} + b_{21}k_{1})\nonumber\\
k_{3} &= hy^{\prime}(x_{n} + a_{3}h,y{n} + b_{31}k_{1} + b_{32}k_{2})\nonumber\\
&\qquad\ldots\nonumber\\
k_{6} &= hy^{\prime}(x_{n} + a_{6}h,y{n} + b_{61}k_{1} + \ldots + b_{65}k_{5}),
\end{align}
$h$ is the step size and the $a_{i}$ and $c_{i}$ coefficients are Cash-Karp parameters, which can be found (along with a more detailed discussion of the iterative method) in Ref.~\cite{NumericalRecipes}.  The embedded fourth order estimate is given by
\begin{equation}
y^{\star}_{n+1} = y_{n} + c^{\star}_{1}k_{1} + c^{\star}_{2}k_{2} + c^{\star}_{3}k_{3} + c^{\star}_{4}k_{4} + c^{\star}_{5}k_{5} + c^{\star}_{6}k_{6} + \mathcal{O}(h^5),
\end{equation}
which leads to an error estimate of
\begin{equation}
y_{n+1} - y^{\star}_{n+1} = \sum_{i=1}^{6}(c_{i} - c^{\star}_{i})k_{i}.
\end{equation}
Once again values used for the Cash-Karp parameters $c^{\star}_{i}$ can be found in Ref.~\cite{NumericalRecipes}.  We select values for the step size that render the discretisation errors negligible.

\section{Simulation Results}
\label{sect:SimulationResults}
We are now in a position to present accurate visualisations of flight trajectories.  The initial simulation parameters are shown in Table \ref{table:InitialValues}.  Two sets of initial conditions are considered in order to better observe the effect boundary layer flow transition has on the flight path. 
%
%
\begin{table}[ht]
\centering
\begin{tabular}{c | c | c}
 \hline
\noalign{\vspace{3pt}}
    Input Variable & Set 1 Value & Set 2 Value  \\
\noalign{\vspace{3pt}}
    \hline 
    \hline
\noalign{\vspace{3pt}}
    Gravitational acceleration $g$ & 9.81ms$^{-2}$ & 9.81ms$^{-2}$\\\hline
    Ball diameter & 0.22m & 0.22m\\\hline 
    Mass of ball & 0.43kg & 0.43kg\\\hline 
    Distance from goal & 25m & 25m\\\hline 
    Distance right of goal center & 0.0m & 0.0m\\\hline 
    Height of ball when kicked & 0.11m (= radius) & 0.11m (= radius)\\\hline 
    Initial speed & 34ms$^{1}$ & 27ms$^{1}$\\\hline
	Initial velocity angle of inclination & 12.5$^{\circ}$ & 18$^{\circ}$\\\hline
	Initial velocity angle of rotation about vertical & 15$^{\circ}$ & 15$^{\circ}$\\\hline
	Ball spin & 6.0rps & 6.0rps\\\hline
	Spin angle of inclination & 90$^{\circ}$ & 90$^{\circ}$\\\hline
	Spin angle of rotation about vertical & 0.0$^{\circ}$ & 0.0$^{\circ}$\\\hline
	Wind speed & 0.0ms$^{-1}$ & 0.0ms$^{-1}$\\\hline
	Angle of wind origin clockwise from straight ahead & 0.0$^{\circ}$ & 0.0$^{\circ}$\\\hline
	Time increment for solver & 0.01s & 0.01s\\\hline
\end{tabular}
\caption{A Table showing the two sets of initial conditions used.}
\label{table:InitialValues}
\end{table}
%
%
We note that the only parameters to change between Set 1 and Set 2 are the initial speed of the ball and the angle of inclination.  Typical free kick type shots are generally taken at $\approx$ 30 $\pm$ 5 ms$^{-1}$~\cite{JEGoff, AsaiTSeoKKobayashiOandSakashitaR, CarreMJAsaiTAkatsukaTandHaakeSJ, BrayKandKerwinDG}, a range in which both initial speeds fall within.  

\begin{figure}[p!] 
  \begin{center} 
  \captionsetup[subfigure]{width=0.42\textwidth}
  \subfloat[]
  {\includegraphics[width=0.46\textwidth]{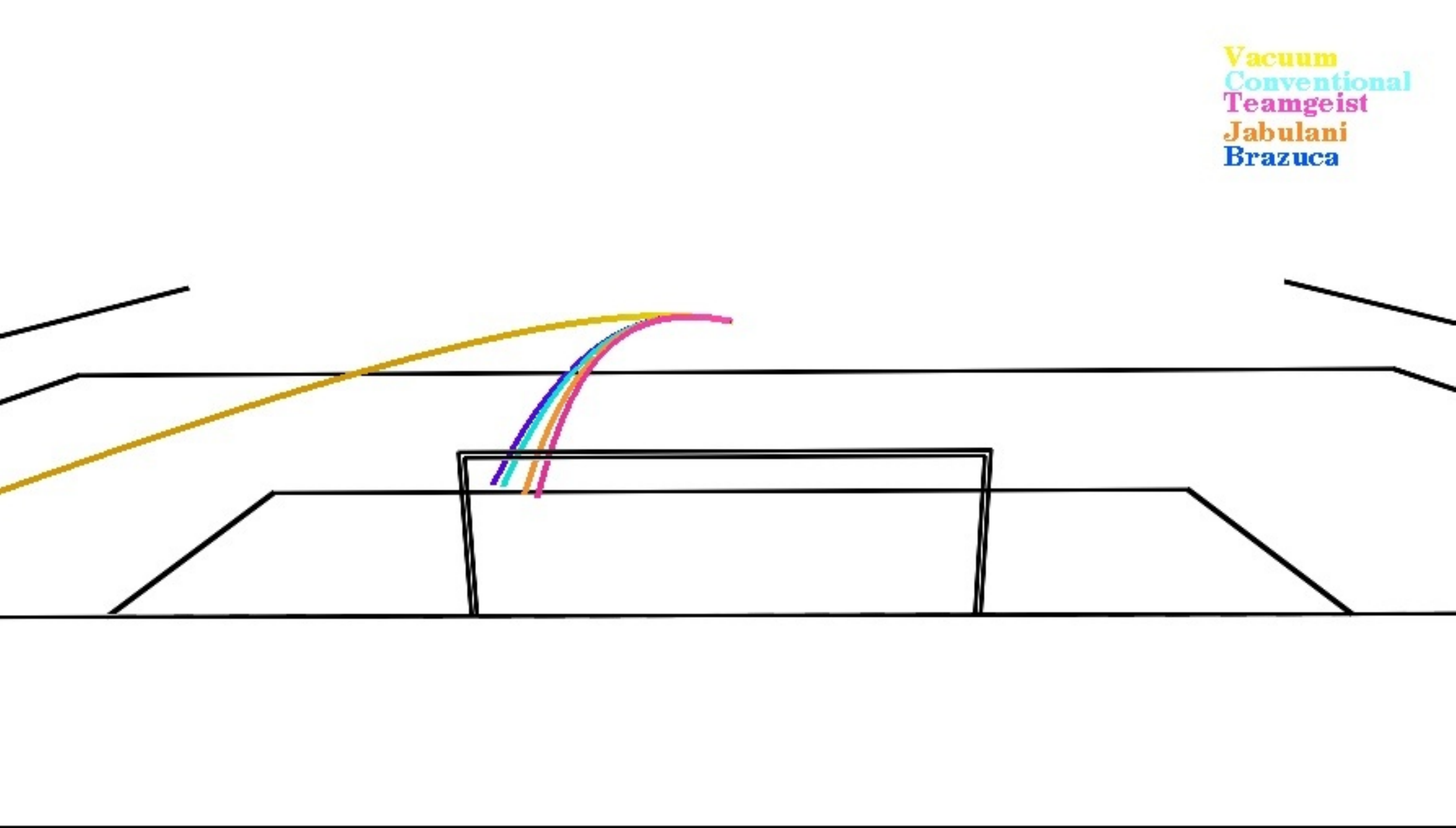}}\,\,
   \subfloat[]
  {\includegraphics[width=0.46\textwidth]{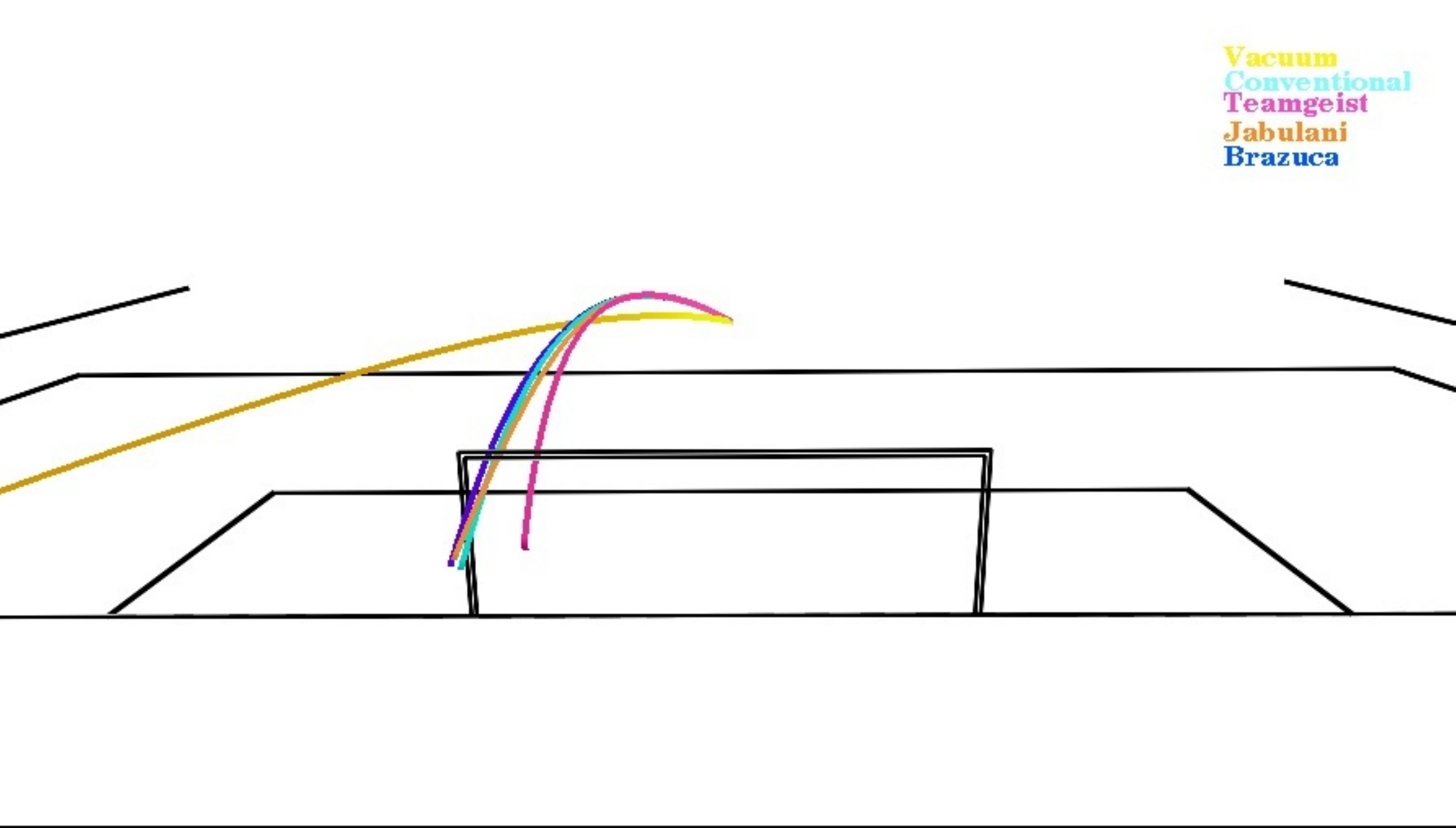}}\\
  \subfloat[]
  {\includegraphics[width=0.46\textwidth]{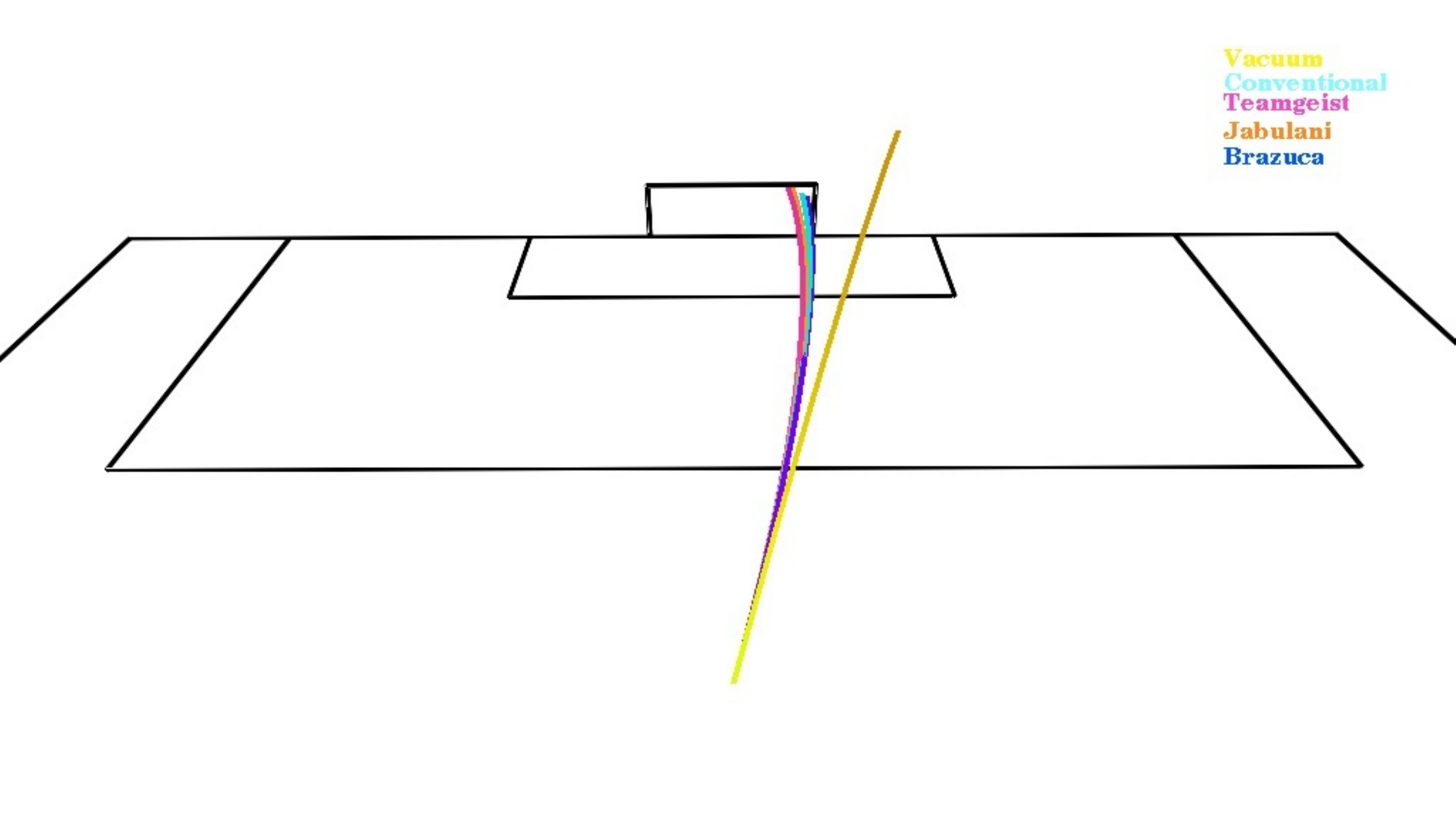}}\,\,
  \subfloat[]
  {\includegraphics[width=0.46\textwidth]{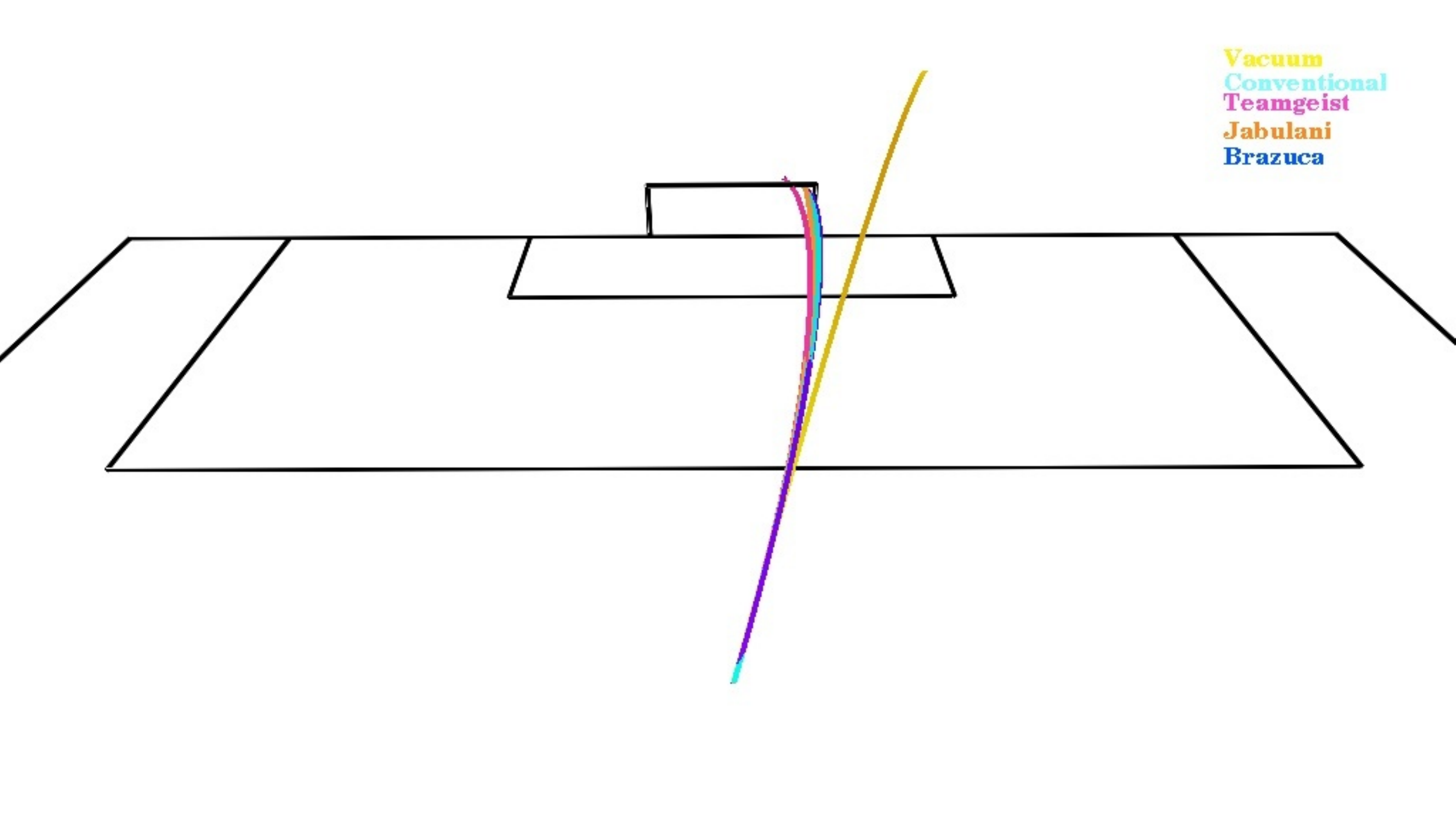}}\\
  \subfloat[]
  {\includegraphics[width=0.46\textwidth]{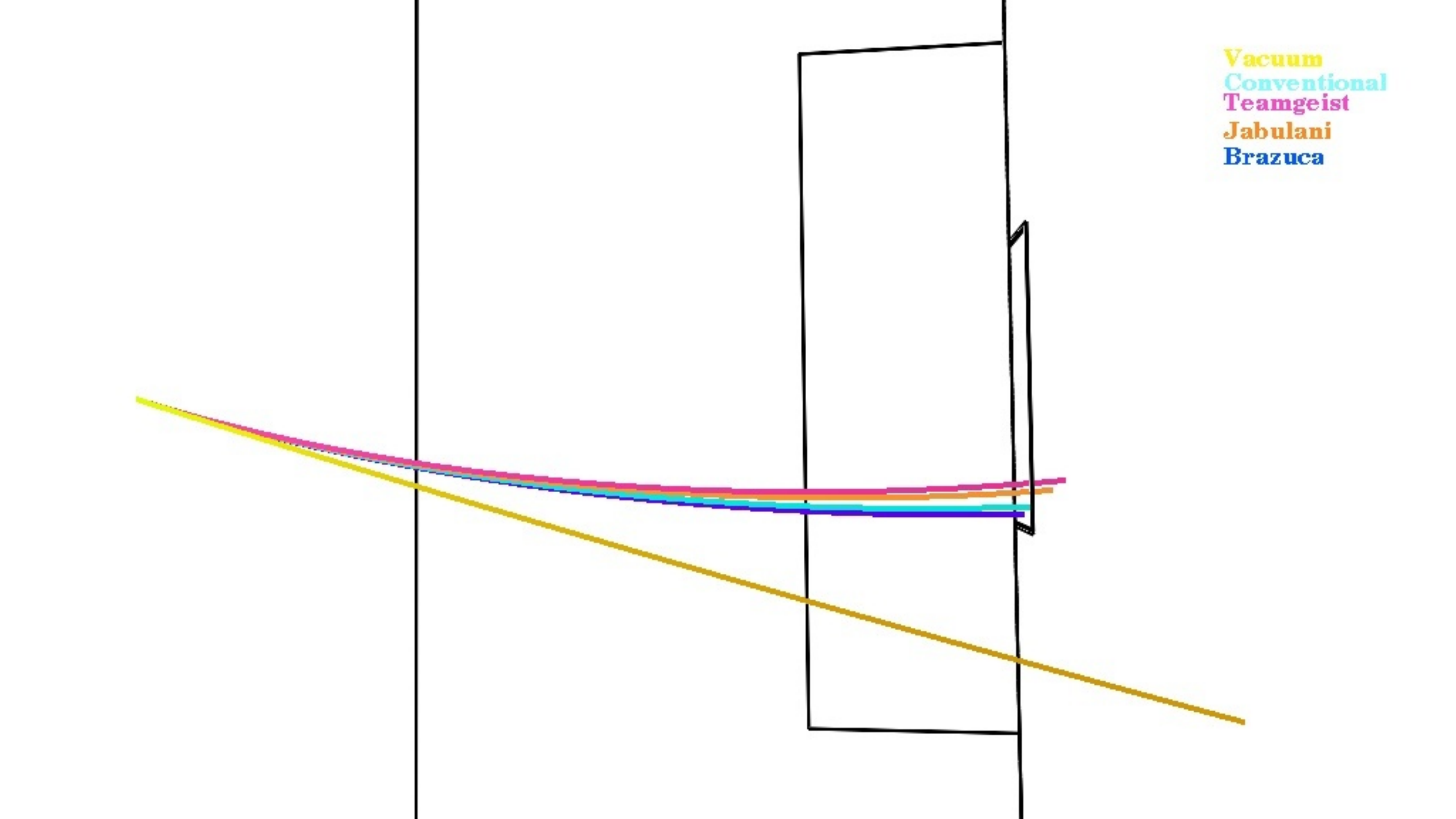}}\,\,
  \subfloat[]
  {\includegraphics[width=0.46\textwidth]{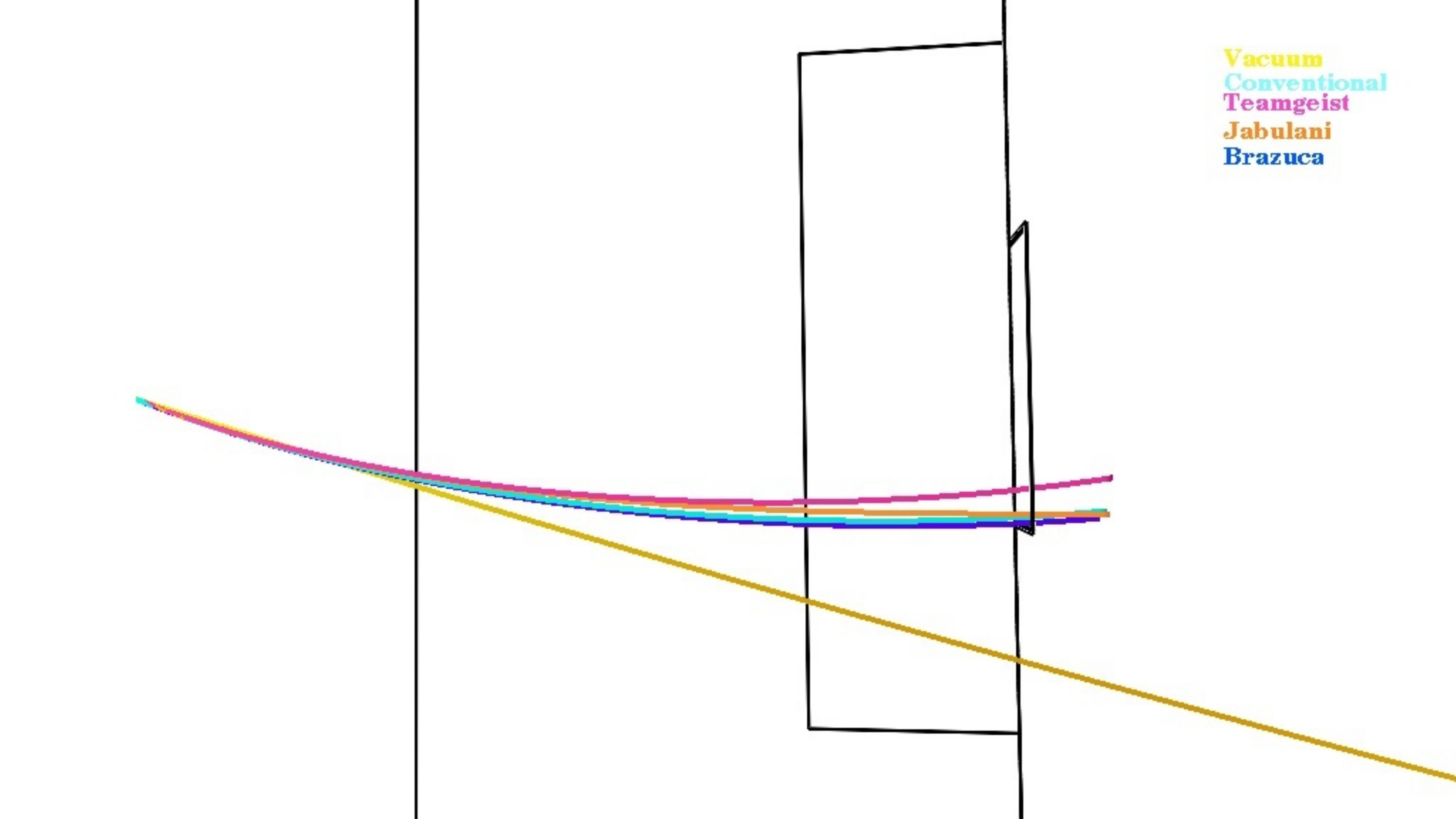}}
 \end{center}
\vspace{-0.5cm}
 \caption{(Colour online) Diagrams depicting the flight paths of various balls.  Subfigures a), c) and e) on the left-hand side correspond to the flight paths obtained using the initial conditions of Set 1 in Table \ref{table:InitialValues}, while subfigures b), d) and f) on the right-hand side correspond to the initial values in Set 2.  The colours of the curves denote the different balls under consideration and the yellow curve illustrates the trajectory of all balls in a vacuum where the lift and drag forces are absent.  Balls include the Brazuca (blue), Conventional 32 Panel ball (aqua), Jabulani (orange) and Teamgeist (purple).}
 \label{fig:AllBallsCompare}
\end{figure}

The illustrations of Figure \ref{fig:AllBallsCompare} show this difference to have a significant impact, particularly on the Jabulani's trajectory.  While the Set 1 initial speed of 34ms$^{-1}$ is sufficiently fast that the boundary layer on all balls considered stays turbulent right until reaching the goal, this is not the case with the initial speed of 27 ms$^{-1}$ used in Set 2.  Recall from Figure \ref{fig:CdFitsCompare} that at approximately 24 ms$^{-1}$ the Jabulani begins its transition from the turbulent to laminar regime, dramatically losing lift as the boundary layer moves away from the ball's surface.  This is particularly evident in the aforementioned figure where the Jabulani has encountered its transition, losing lift and crossing over the trajectory of the conventional 32-panel ball.  This sensitivity of the lift coefficient to perturbation of initial speed within the speed range typical shots may arrive at the keeper, can further add to the inherent difficulty of the knuckleball effects the Jabulani experiences, discussed in Section \ref{sect:DragCoefficients}.
The angle of inclination is altered between the two sets of initial conditions in order for the ball to reach the goalmouth in a comparable position for both sets.  Naturally, as players are aiming for the goal, this reflects gameplay.  Figure \ref{fig:AllBallsCompare} also highlights the Teamgeist as the most interesting ball from an aerodynamics point of view.  The long deep seams keep the flight in the turbulent regime, while its smooth panels keep the boundary layer intact at the highest speeds maximising lift and associated curvature in the trajectory.

\begin{figure}[p!] 
  \begin{center} 
  \captionsetup[subfigure]{width=0.42\textwidth}
  \subfloat[]
  {\includegraphics[width=0.46\textwidth]{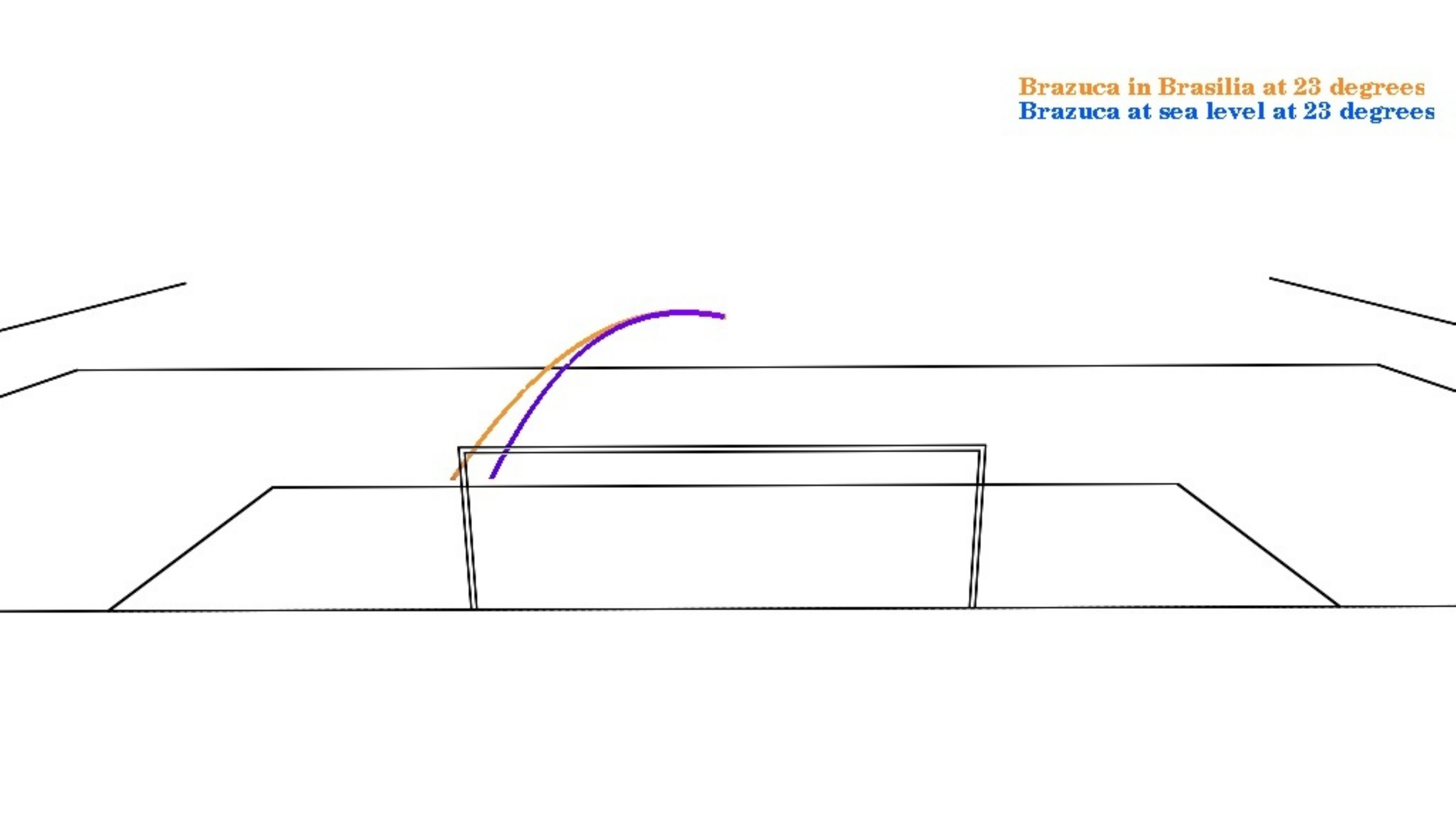}}\,\,
   \subfloat[]
  {\includegraphics[width=0.46\textwidth]{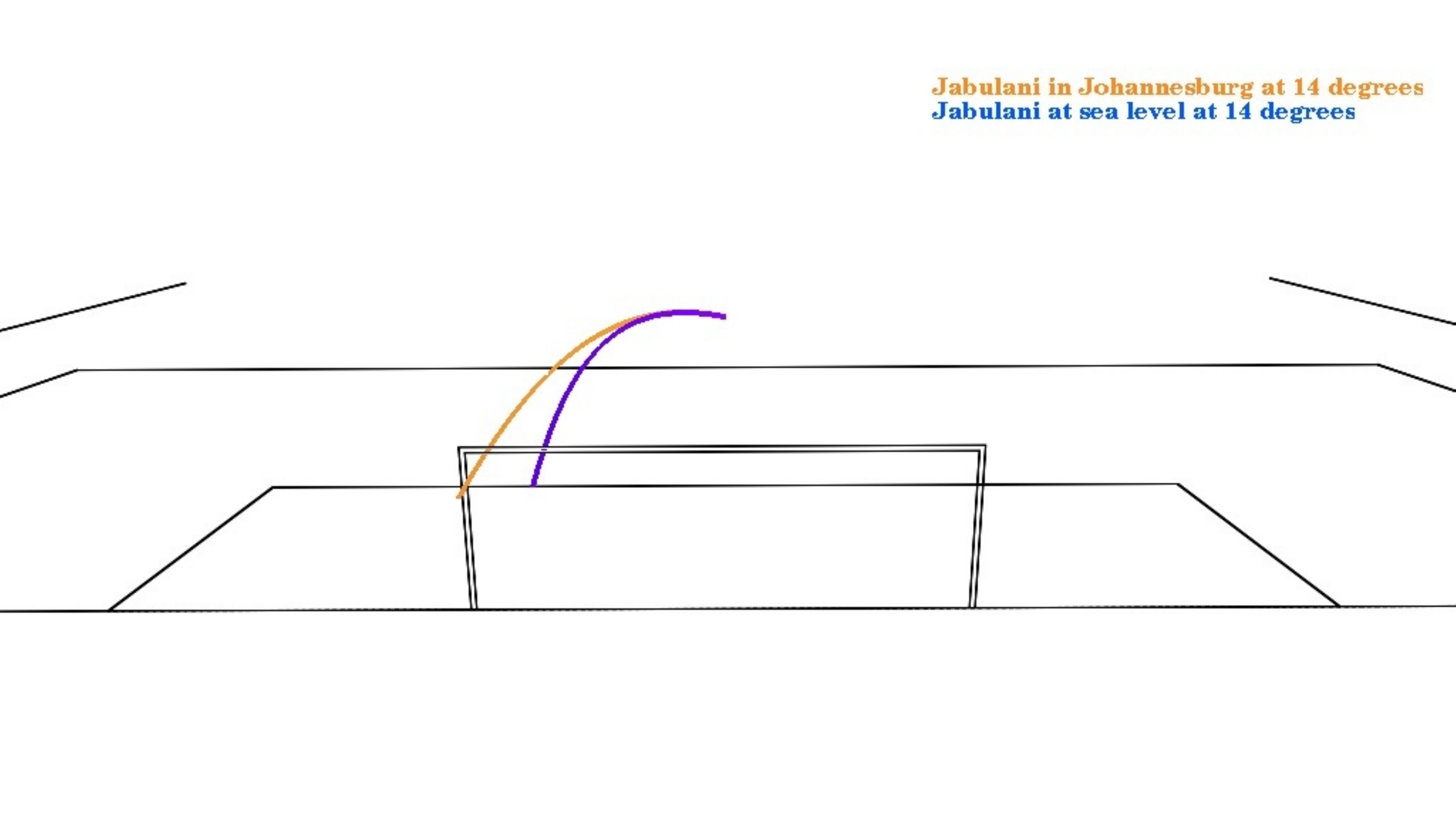}}\\
  \subfloat[]
  {\includegraphics[width=0.46\textwidth]{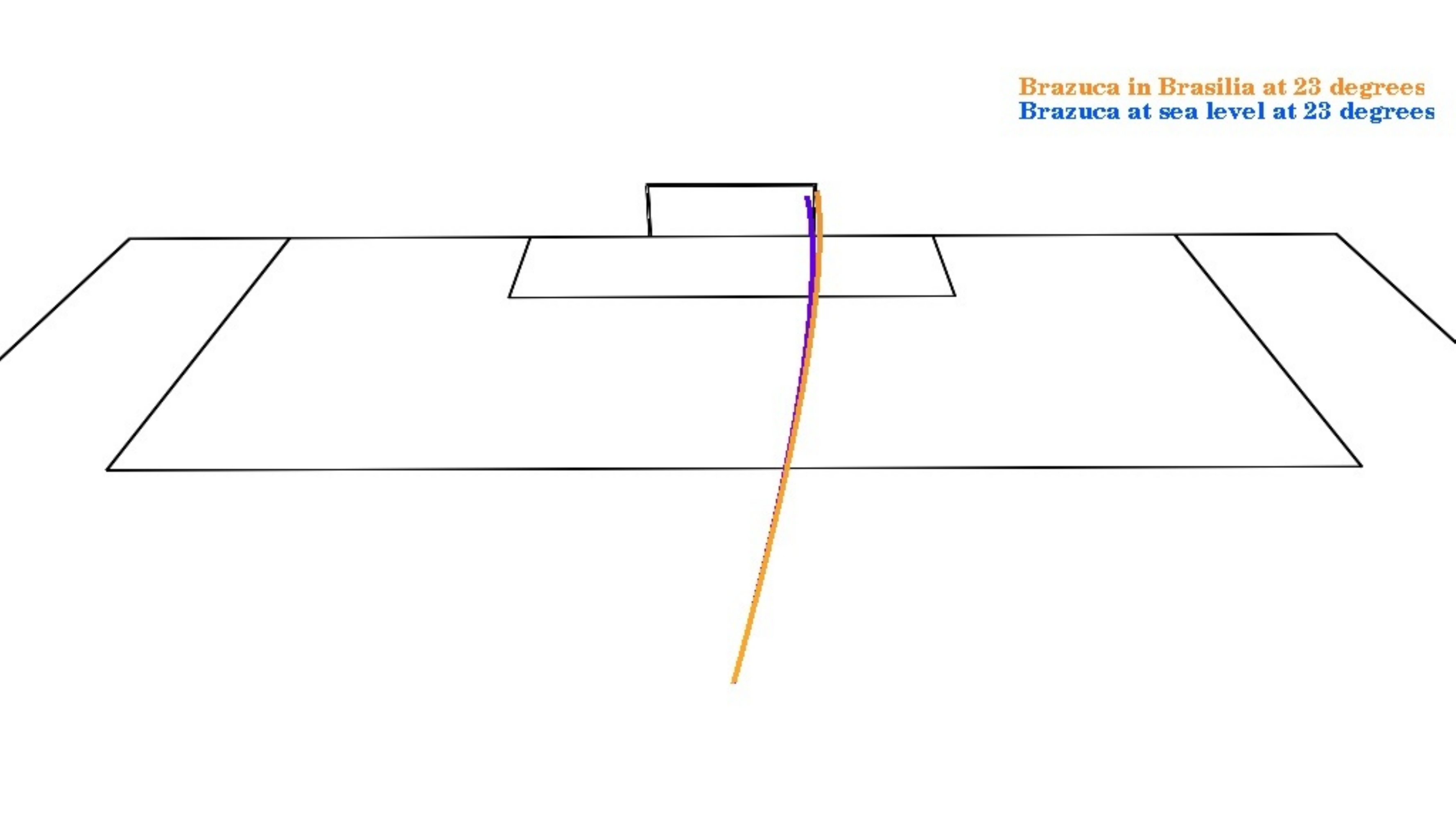}}\,\,
  \subfloat[]
  {\includegraphics[width=0.46\textwidth]{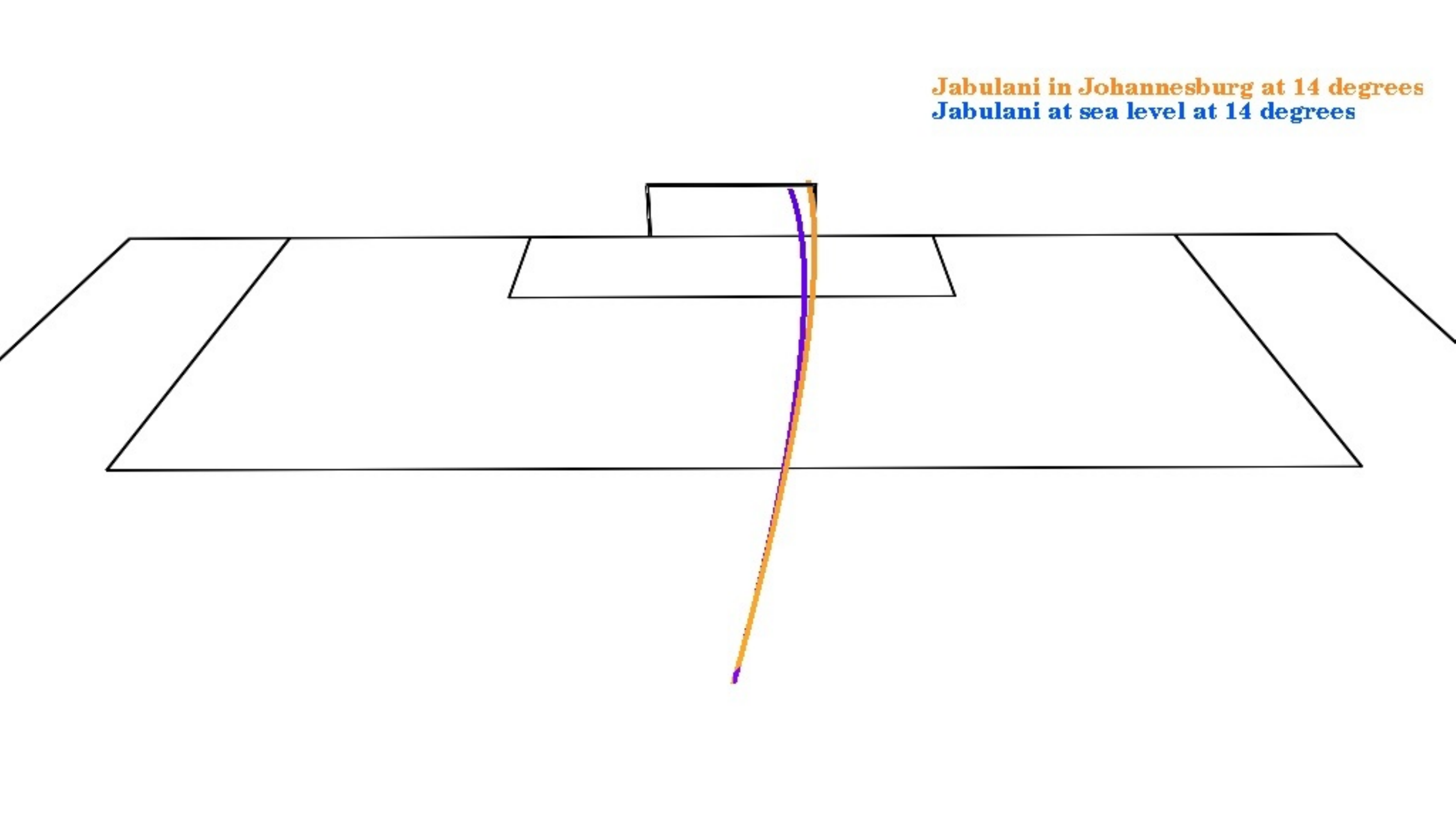}}\\
  \subfloat[]
  {\includegraphics[width=0.46\textwidth]{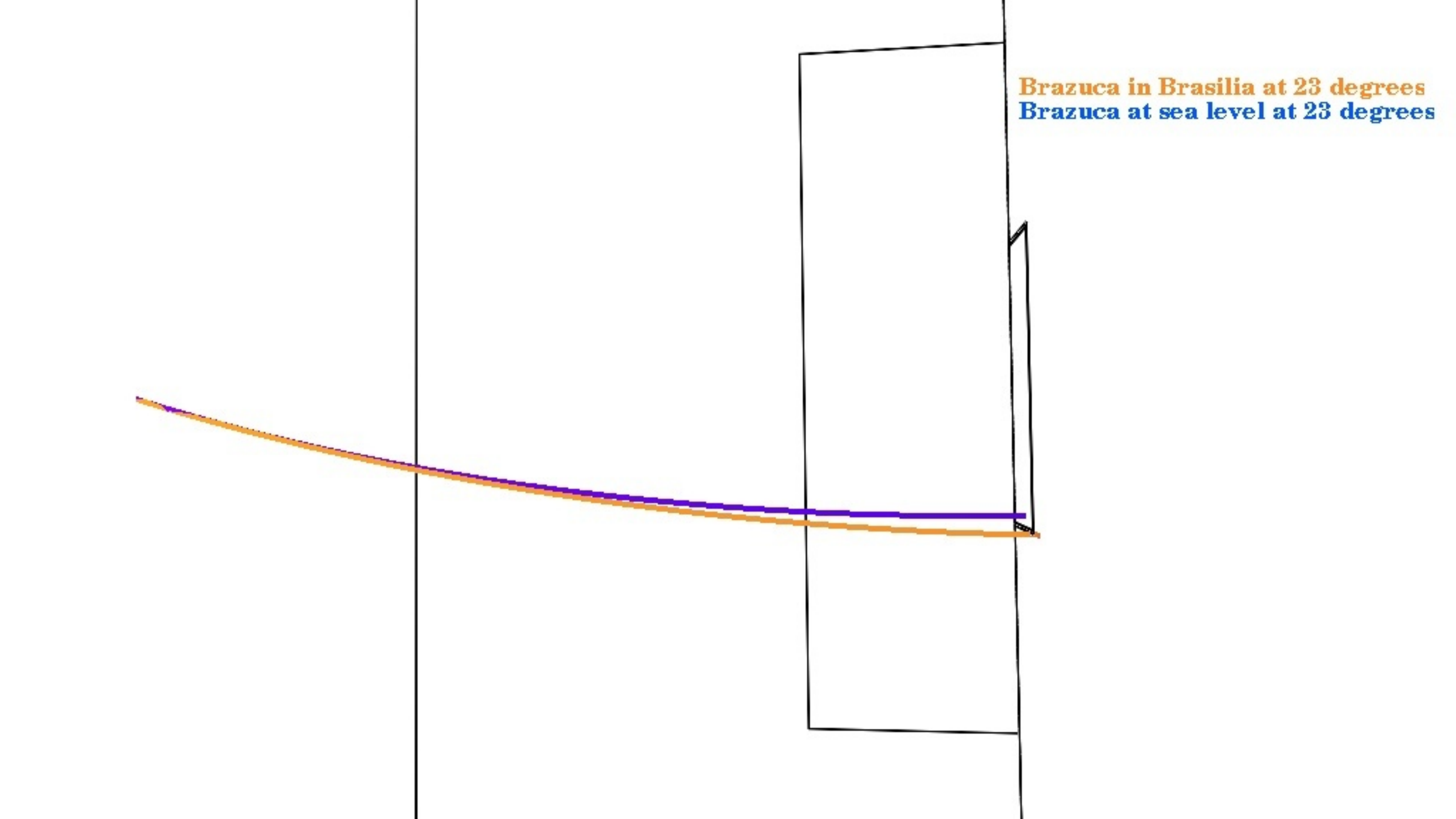}}\,\,
  \subfloat[]
  {\includegraphics[width=0.46\textwidth, trim=0cm 0.15cm 0cm 0cm, clip]{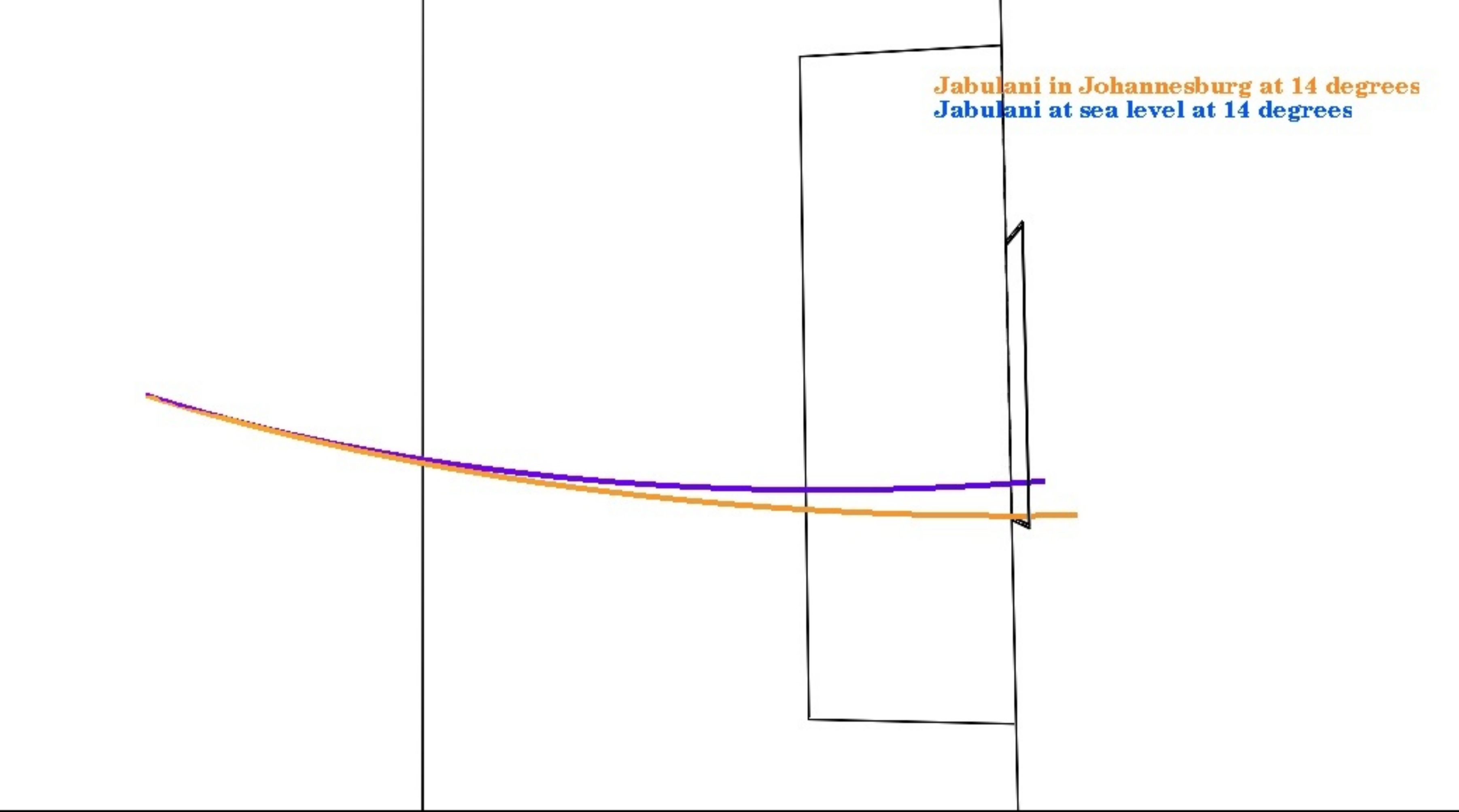}}
 \end{center}
\vspace{-0.5cm}
 \caption{(Colour online) Diagrams comparing the flight paths of the Brazuca and Jabulani over the range of altitudes encountered at their prospective world cups.  Subfigures a), c) and e) on the left-hand side show comparisons with the Brazuca, while subfigures b), d) and f) on the right hand side depict Jabulani flight paths.  The temperatures of 23$^{\circ}$C and 14$^{\circ}$C were the recorded temperatures at the respective final's kickoff time.}
 \label{fig:AltComp}
\end{figure}

Figure \ref{fig:AltComp} shows a comparison of the Brazuca and Jabulani with changing altitude.  The initial conditions used for these simulations are Set 1 of Table \ref{table:InitialValues}.  As discussed in Section \ref{sect:Altitude} the capital city in which the last two world cup finals were played was the venue with the highest elevation, while both world cups hosted matches played at sea level.  The temperatures of 23$^{\circ}$C and 14$^{\circ}$C were chosen for the simulation as these were the pitchside temperatures at kickoff of the respective world cup finals~\cite{2014FinalInfoFIFA}.  We readily observe that changing the altitude at which one plays can significantly alter flight trajectories.  At the time the ball went out of play (either for a goalkick or into the net), the difference in position for the Brazuca sea level kick and the one in Brasilia was 49.0 cm, while the corresponding difference for the Jabulani was notably more, at 82.0 cm.  This relatively large variation in the Jabulani's behaviour derived solely from the different location at which the kick was taken may well have further compounded previously discussed difficulties inherent to the ball.

\begin{figure}[p!] 
  \begin{center} 
  \captionsetup[subfigure]{width=0.42\textwidth}
  \subfloat[]
  {\includegraphics[width=0.46\textwidth]{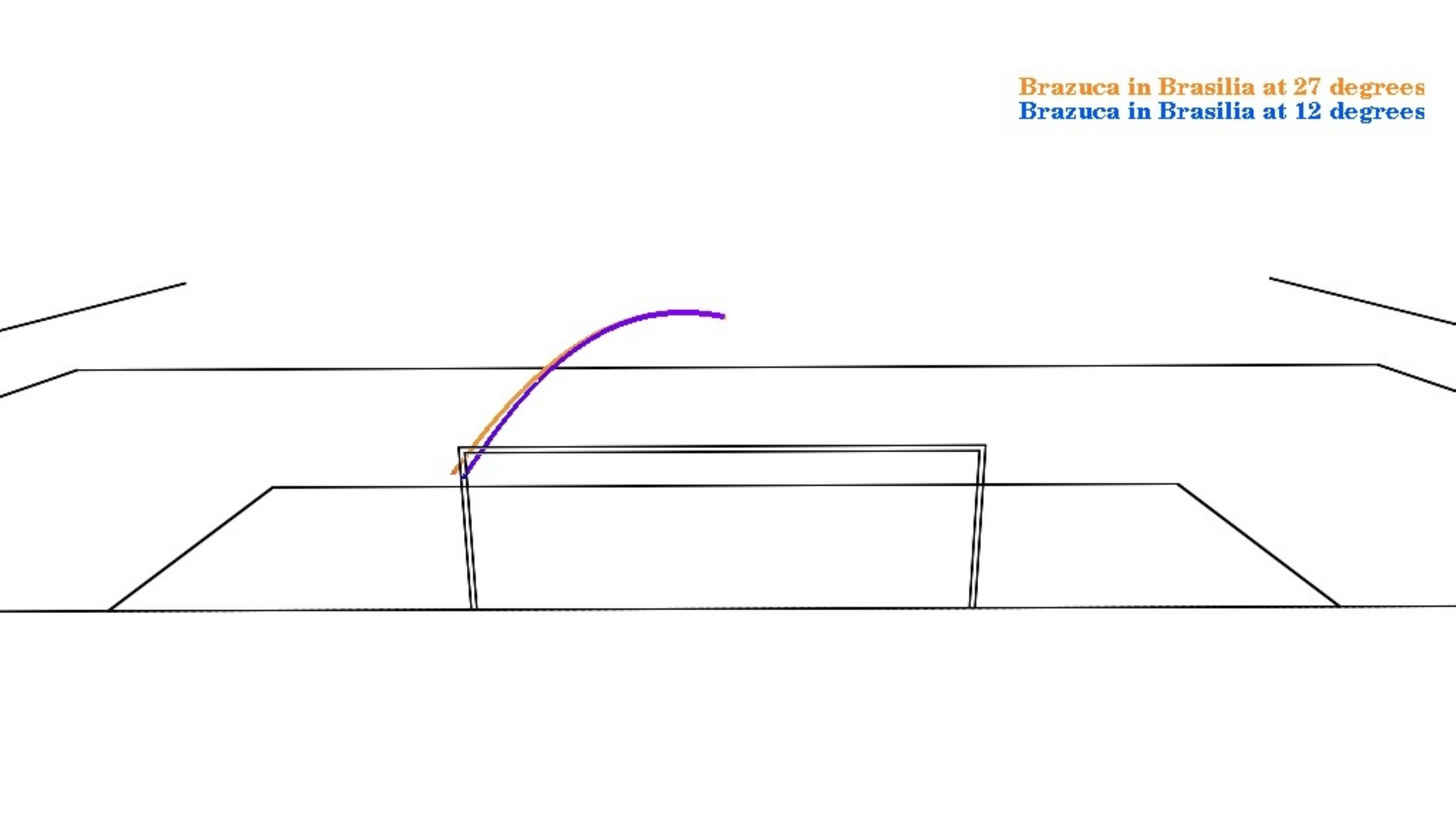}}\,\,
   \subfloat[]
  {\includegraphics[width=0.46\textwidth]{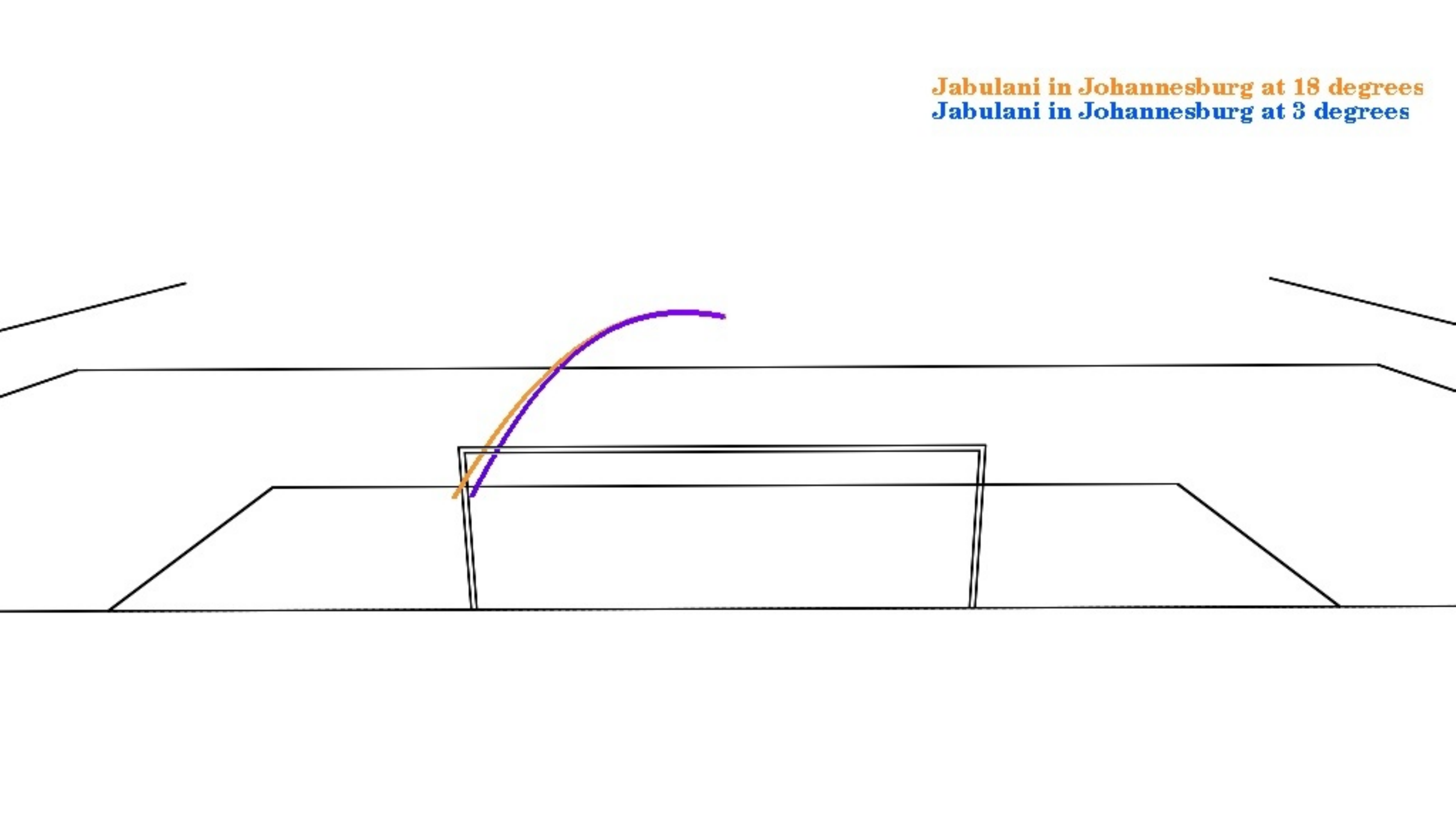}}\\
  \subfloat[]
  {\includegraphics[width=0.46\textwidth]{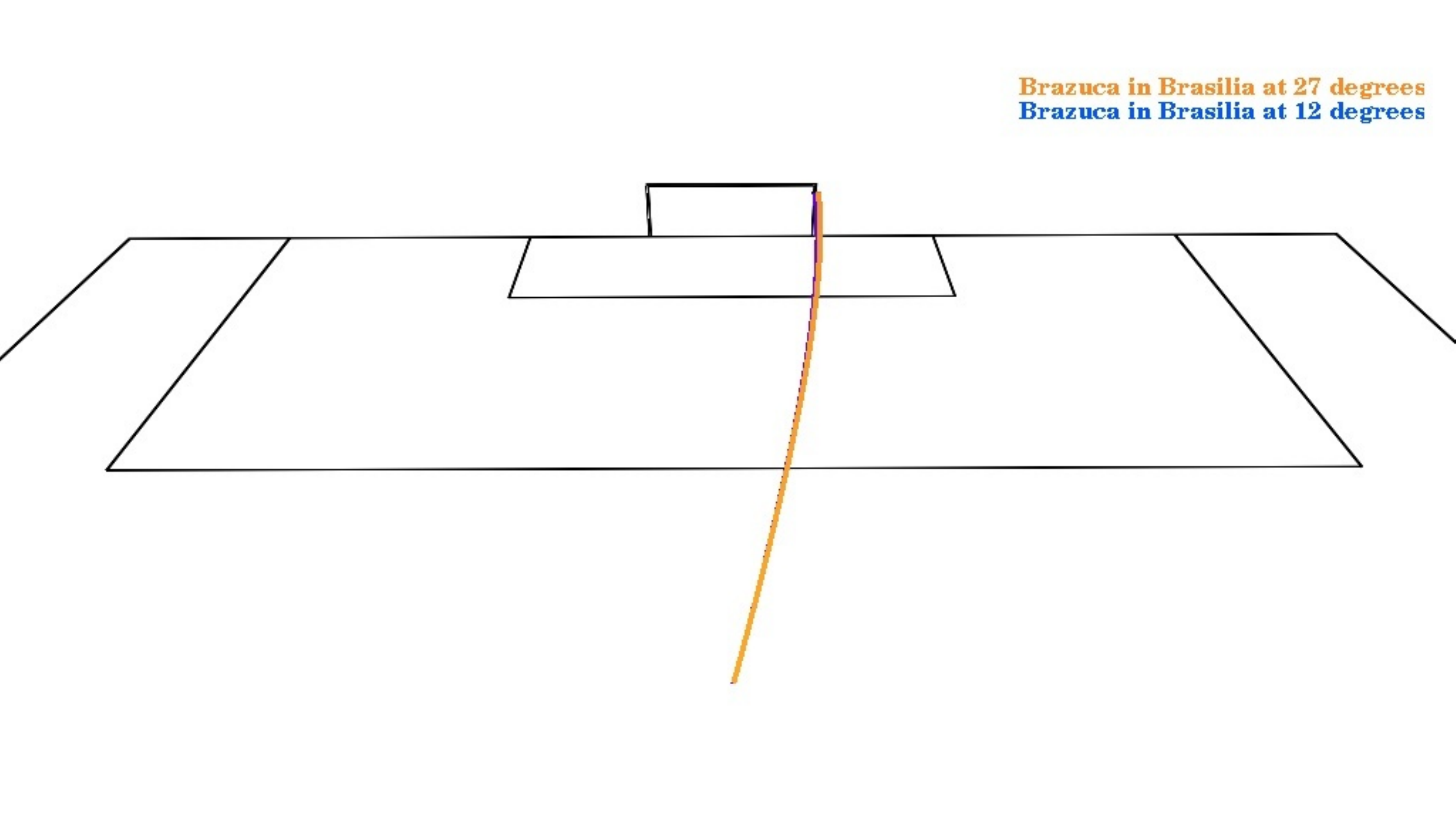}}\,\,
  \subfloat[]
  {\includegraphics[width=0.46\textwidth]{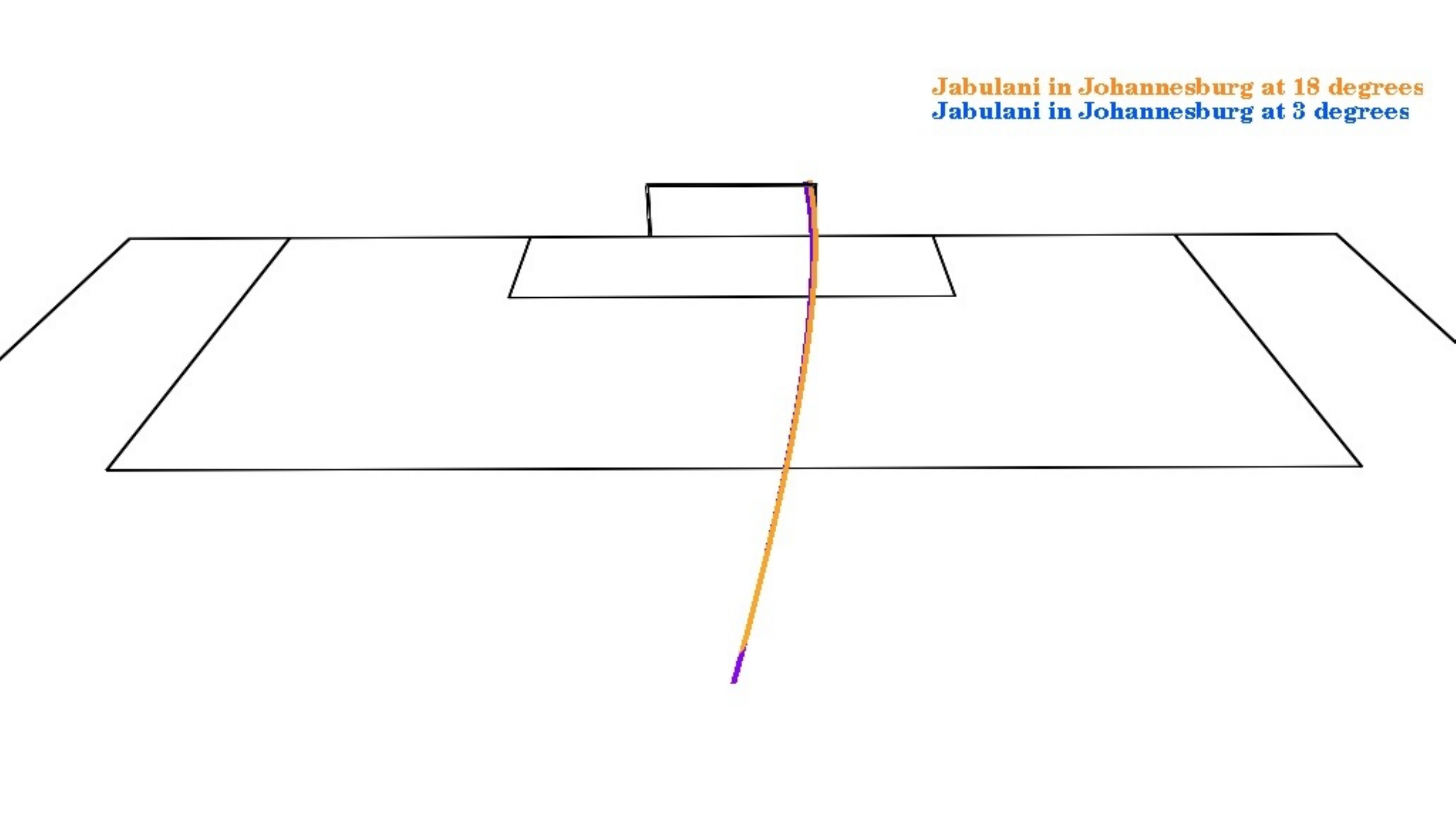}}\\
  \subfloat[]
  {\includegraphics[width=0.46\textwidth]{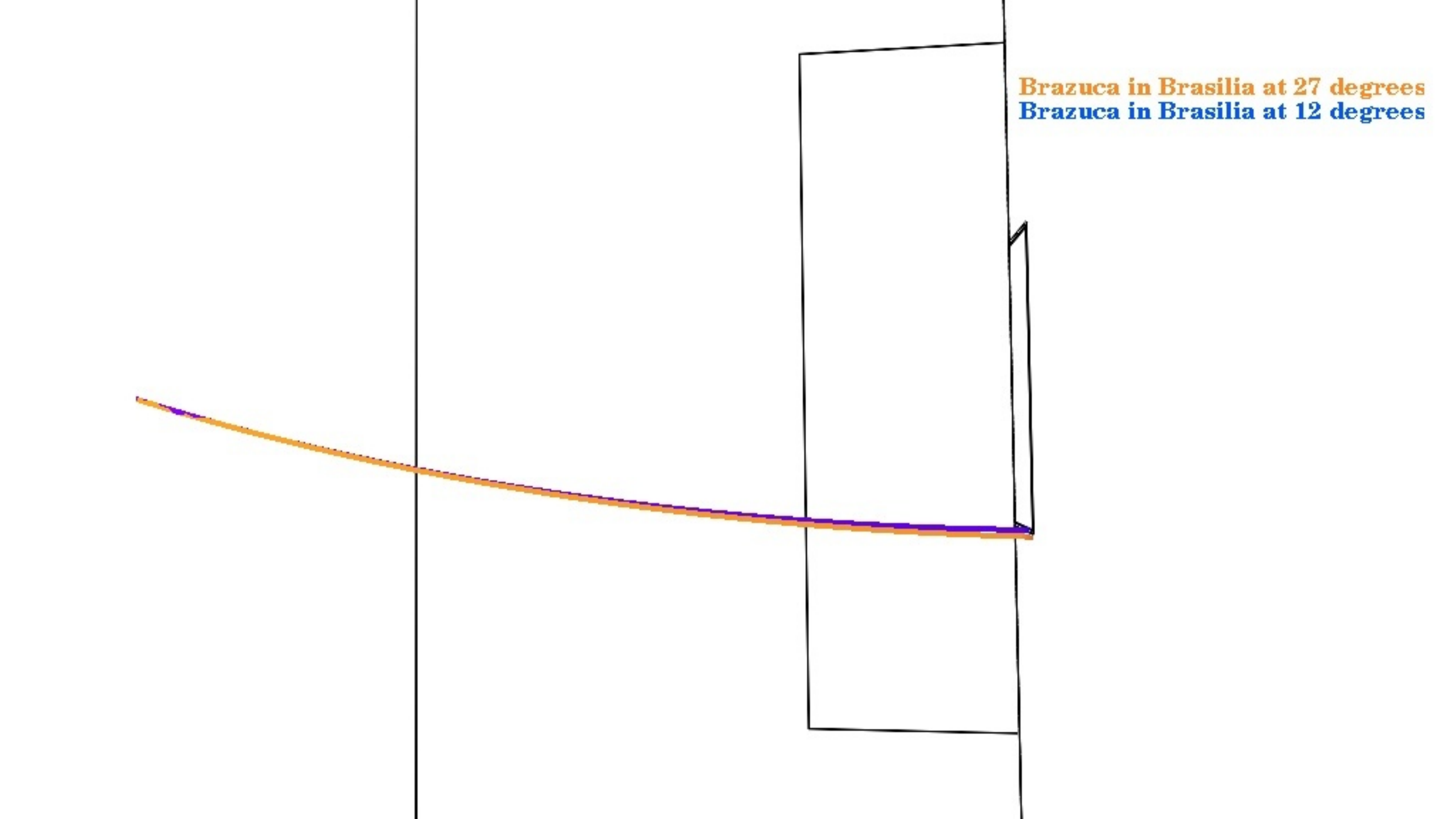}}\,\,
  \subfloat[]
  {\includegraphics[width=0.46\textwidth]{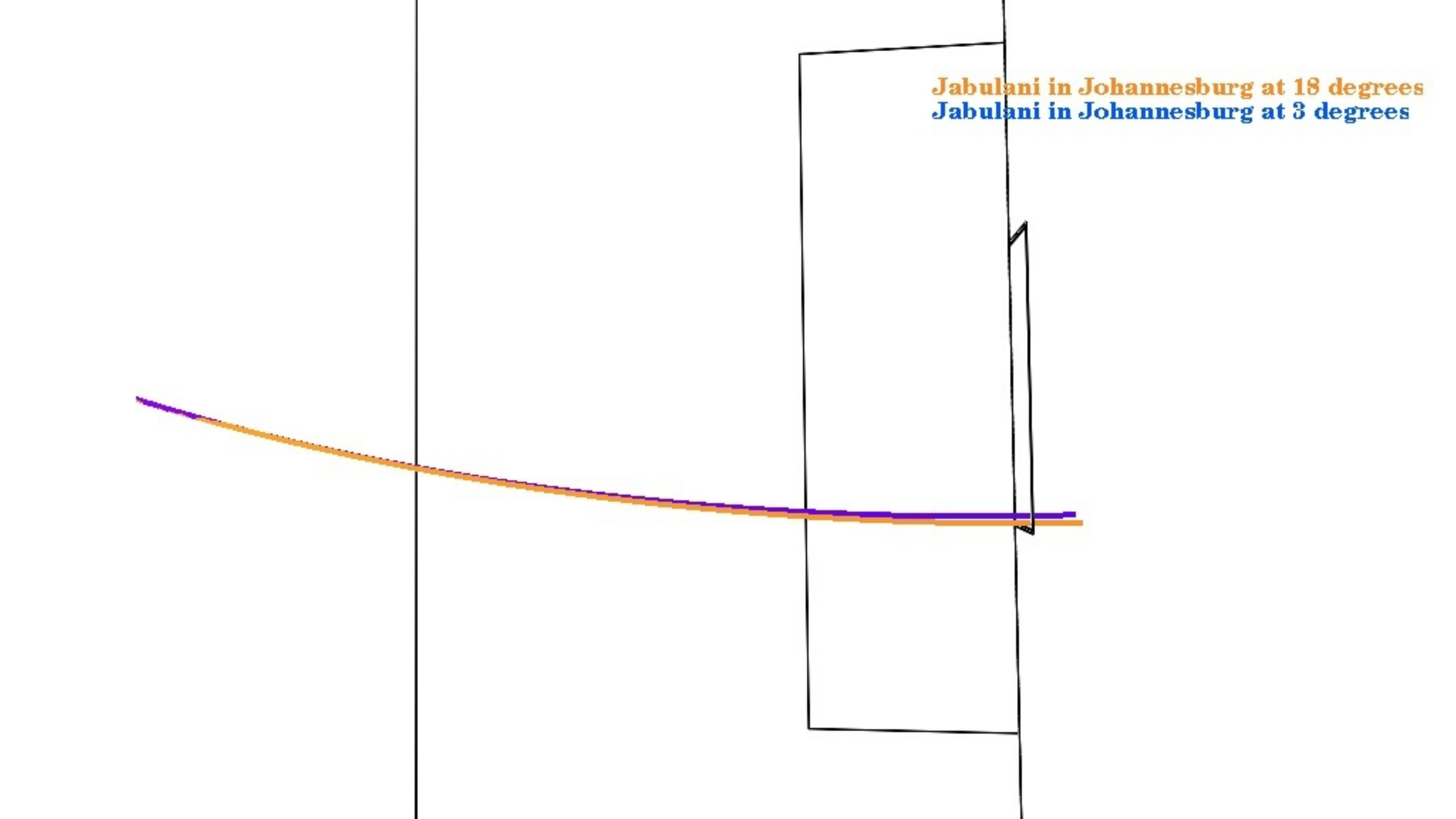}}
 \end{center}
\vspace{-0.5cm}
 \caption{(Colour online) Diagrams comparing the flight paths of the Brazuca and Jabulani over the range of temperatures likely encountered at their prospective world cups.  Subfigures a), c) and e) on the left hand side show comparisons with the Brazuca, while subfigures b), d) and f) on the right hand side depict Jabulani flight paths.  The temperatures of 12$^{\circ}$C and 27$^{\circ}$C were the average minimum and maximum temperatures during the tournament months of June and July in Bras\'ilia, while 3$^{\circ}$C and 18$^{\circ}$C were the average minimum and maximum temperatures during June and July in Johannesburg.}
 \label{fig:TempComp}
\end{figure}

While the altitude variation was significant, the difference in flight trajectories caused by the temperature variation effect on air density is not as great, as can be readily deduced from Figure \ref{fig:AirDensityvsAltitude}.  Figure \ref{fig:TempComp} compares two kicks at the average minimum and maximum temperatures during the tournament months of June and July for both the Brazuca and Jabulani in the left hand and right hand columns respectively.  At the time the ball went out of play (either for a goalkick or into the net), the difference in position for the Brazuca at 12$^{\circ}$C and the Brazuca at 27$^{\circ}$C was 16.6 cm while the corresponding distance between the Jabulani kicked at 18$^{\circ}$C and 3$^{\circ}$C was 24.5 cm, both of which are significantly less than the variation in trajectory seen resulting from changes in altitude.  

%
\section{Conclusions}
\label{sect:Conclusions}
Following a general summary of ball aerodynamics, we have developed a method for predicting the behaviour of lift coefficients based on drag coefficient data, lift coefficient data as available and an understanding of the physics of the boundary layer.  The drag coefficient data has also been fitted with an improved functional form addressing high flow rate effects.  This has been utilised to run accurate simulations of various balls over typical match parameters.

While it was known that the high transition speed of the Jabulani caused knuckleball type effects at higher speeds, our boundary layer discussion indicates that this transition is likely to coincide with a loss of lift.  This has the potential to further compound other difficulties with the ball, leading us to conclude that the players complaints in the 2010 world cup is supported by the available data and the present analysis.

We then considered the effect altitude and temperature has on flight trajectories, quantifying the difference in flight paths for the Brazuca and Jabulani at various altitudes and temperatures experienced at their respective world cups.  We found that the variation observed in the Jabulani's flight path in South Africa due to altitude variation was significant, perhaps further adding to the difficulty the players experienced during the 2010 FIFA world cup.

While predicted lift coefficients from drag data coupled with an understanding of the boundary layer provides an attractive framework within which we can produce accurate simulations, empirical lift data would be a natural extension.
It would therefore be interesting to confront our theoretical predictions summarised in Equation \ref{eq:CLPrediction} with experimentally measured lift coefficient values over a range of Reynolds numbers and spin parameters for a variety of balls, whether they be wind tunnel data or the results of trajectory analysis.

\section*{Acknowledgments}
DBL thanks Jamie Seidel and the Advertiser newspaper for their interest and enthusiasm for exploring the physics of sports ball aerodynamics.
The authors also thank Michels Warren for their permission to include the diagrams illustrating boundary layer separation in Section \ref{sect:FundamentalBallAerodynamics}.  Their generosity is greatly appreciated.  This research was supported by the School of Physical Sciences at the University of Adelaide.

%
%
%
%
%
%

\end{document}